\pdfoutput=1%
\documentclass[10pt,twoside]{article}
\usepackage{amsmath,amsfonts,amssymb}
\usepackage{balance}
\usepackage[format=plain,justification=raggedright,singlelinecheck=false,font=small,labelfont=bf,labelsep=space]{caption}
\usepackage{fancyhdr}
\usepackage[T1]{fontenc}
\usepackage[papersize={15.6cm,23.4cm},text={11.3cm,20.4cm},centering]{geometry}
\usepackage{graphicx}
\usepackage[utf8]{inputenc}
\usepackage{lastpage}
\usepackage[comma,sort&compress,super]{natbib}
\usepackage{sectsty}
\usepackage{times,txfonts}
\usepackage[textsize=footnotesize]{todonotes}
\usepackage{xspace}
\usepackage{hyperref}
\usepackage{hypernat}

\usepackage{endnotes}
\newlength{\figwidth}
\newlength{\figwidthsmall}
\newlength{\figwidthfull}
\newcommand{\chisqr}{\ensuremath{\chi^2}}
\newcommand{\cost}{\ensuremath{\langle\cos^2\theta_\text{2D}\rangle}}

\newcommand{\degree}{\ensuremath{^\circ}}%
\newcommand{\DIBNformula}{C$_{7}$H$_{3}$I$_{2}$N}%

\newcommand{\eg}{e.\,g.}%
\newcommand{\etal}{et al.}%

\newcommand{\ie}{i.\,e.}%
\newcommand{\Iplus}{\ensuremath{\text{I}^+}\xspace}%
\newcommand{\Iplusn}{\ensuremath{\text{I}^{+n}}\xspace}%
\newcommand{\Iplusone}{\ensuremath{\text{I}^{+1}}}%
\newcommand{\Iplusseven}{\ensuremath{\text{I}^{+7}}}%
\newcommand{\IYAG}{\ensuremath{\text{I}_\text{YAG}}\xspace}%
\newcommand{\INoYAG}{\ensuremath{\text{I}_\text{NoYAG}}\xspace}%
\newcommand{\IYAGmNoYAG}{\ensuremath{\IYAG-\INoYAG}\xspace}%
\newcommand{\liod}{\ensuremath{d}}
\newcommand{\M}{\ensuremath{\text{M}}\xspace}%
\newcommand{\Sii}{\ensuremath{S(\liod)}}%
\newcommand{\thetadd}{\ensuremath{\theta_\text{2D}}}%
\newcommand{\twotheta}{\ensuremath{2\Theta}}%

\fancypagestyle{plain}{
\fancyhead[L]{\includegraphics[height=8pt]{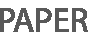}}
\fancyhead[C]{\hspace{-1cm}\includegraphics[height=15pt]{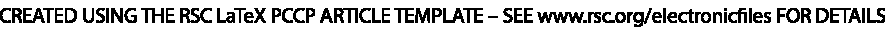}}
\fancyhead[R]{\includegraphics[height=10pt]{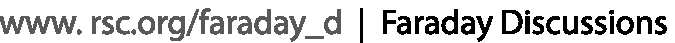}\vspace{-0.2cm}}
}

\makeatletter
\renewcommand{\fnum@figure}{\textbf{Fig.~\thefigure~~}}
\def\subsubsection{\@startsection{subsubsection}{3}{10pt}{-1.25ex plus -1ex minus -.1ex}{0ex plus 0ex}{\normalsize\bf}}
\def\paragraph{\@startsection{paragraph}{4}{10pt}{-1.25ex plus -1ex minus -.1ex}{0ex plus 0ex}{\normalsize\textit}}
\renewcommand\@biblabel[1]{#1}
\renewcommand\@makefntext[1]%
{\noindent\makebox[0pt][r]{\@thefnmark\,}#1}
\makeatother
\sectionfont{\large}
\subsectionfont{\normalsize}
\makeatletter
\newcommand*{\onlinecite}[2][]{%
  \begingroup
  \let\NAT@mbox=\mbox
  \let\@cite\NAT@citenum
  \let\NAT@space\NAT@spacechar
  \let\NAT@super@kern\relax
  \renewcommand\NAT@open{}%
  \renewcommand\NAT@close{}%
  \cite[#1]{#2}%
  \endgroup
}
\makeatother
\par
\begin{document}
\thispagestyle{plain}
\noindent\sloppy{\LARGE\textbf{%
      Toward atomic resolution diffractive imaging of isolated molecules with x-ray free-electron lasers
   }\par}%
\vspace{0.25cm}%
\noindent{\large\textbf{%
      Stephan Stern,$^{ab}$ %
      Lotte Holmegaard,$\!^{ae}$ %
      Frank Filsinger,$\!^{fg,1}$ %
      Arnaud Rouzée,$\!^{hi}$ %
      Artem Rudenko,$\!^{gjk}$ %
      Per Johnsson,$\!^{l}$ %
      Andrew V. Martin,$\!^{a,2}$ %
      Anton Barty,$\!^{a}$ %
      Christoph Bostedt,$\!^{m}$ %
      John Bozek,$\!^{m}$ %
      Ryan Coffee,$\!^{m}$ %
      Sascha Epp,$\!^{gj}$ %
      Benjamin Erk,$\!^{cgj}$ %
      Lutz Foucar,$\!^{gn}$ %
      Robert Hartmann,$\!^{o}$ %
      Nils Kimmel,$\!^{p}$ %
      Kai-Uwe Kühnel,$\!^{j}$ %
      Jochen Maurer,$\!^{e}$ %
      Marc Messerschmidt,$\!^{m}$ %
      Benedikt Rudek,$\!^{gj,3}$ %
      Dmitri Starodub,$\!^{q,4}$ %
      Jan Thøgersen,$\!^{e}$ %
      Georg Weidenspointner,$\!^{pr,5}$ %
      Thomas A. White,$\!^{a}$ %
      Henrik Stapelfeldt,$\!^{es}$ %
      Daniel Rolles,$\!^{cgn}$ %
      Henry N. Chapman,$\!^{abc}$ %
      and Jochen Küpper$^{abc\ast{}}$ %
   }\par}%
\vspace{0.2cm}

\noindent{\small\textbf{\textit{%
         Received Xth XXXXXXXXXX 2014, \\
         Accepted Xth XXXXXXXXX 2014 \\
         First published on the web Xth XXXXXXXXXX 2014 \\
      }%
      DOI: 10.1039/c0xxxxx}
   \par}
\vspace{0.2cm}
\noindent%
We give a detailed account of the theoretical analysis and the experimental results of an
x-ray-diffraction experiment on quantum-state selected and strongly laser-aligned gas-phase
ensembles of the prototypical large asymmetric rotor molecule 2,5-diiodobenzonitrile, performed at
the Linac Coherent Light Source [\textit{Phys.~Rev.~Lett.} \textbf{112}, 083002 (2014)]. This
experiment is the first step toward coherent diffractive imaging of structures and structural
dynamics of isolated molecules at atomic resolution, \ie, picometers and femtoseconds, using x-ray
free-electron lasers. \footnotetext{\textit{\sloppy
      $^a$ Center for Free-Electron Laser Science (CFEL), Deutsches Elektronen-Synchrotron (DESY),
      Notkestrasse 85, 22607 Hamburg, Germany. \\
      E-mail:~\href{mailto:jochen.kuepper@cfel.de}{jochen.kuepper@cfel.de},
      URL:~\href{http://desy.cfel.de/cid/cmi}{http://desy.cfel.de/cid/cmi}
      \\
      $^b$ Department of Physics, University of Hamburg, Luruper Chaussee 149, 22761 Hamburg,
      Germany.
      \\
      $^c$ Deutsches Elektronen-Synchrotron (DESY), Notkestrasse 85, 22607 Hamburg, Germany.
      \\
      $^d$ The Hamburg Center for Ultrafast Imaging, University of Hamburg, Luruper Chaussee 149,
      22761 Hamburg, Germany.
      \\
      $^e$ Aarhus University, Department of Chemistry, 8000 Aarhus C, Denmark.
      \\
      $^f$ Fritz Haber Institute of the MPG, Faradayweg 4--6, 14195 Berlin, Germany.
      \\
      $^g$ Max Planck Advanced Study Group at CFEL, Notkestrasse 85, 22607 Hamburg, Germany.
      \\
      $^h$ FOM Institute AMOLF, Science Park 104, 1098 XG Amsterdam, The Netherlands.
      \\
      $^i$ Max-Born-Institute, Max Born Str.\ 2a, 12489 Berlin, Germany.
      \\
      $^j$ Max Planck Institute for Nuclear Physics, 69117 Heidelberg, Germany.
      \\
      $^k$ J.\,R.\ Macdonald Laboratory, Department of Physics, Kansas State University, Manhattan,
      KS 66506, USA.
      \\
      $^l$ Lund University, Department of Physics, P.\,O.\ Box 118, 22100 Lund, Sweden.
      \\
      $^m$ Linac Coherent Light Source, SLAC National Accelerator Laboratory, 2575 Sand Hill Road,
      Menlo Park, CA 94025, USA.
      \\
      $^n$ Max Planck Institute for Medical Research, 69120 Heidelberg, Germany.
      \\
      $^o$ PNSensor GmbH, 81739 Munich, Germany.
      \\
      $^p$ Max Planck Semiconductor Laboratory, 81739 Munich, Germany.
      \\
      $^q$ Department of Physics, Arizona State University, Tempe, AZ 85287, USA.
      \\
      $^r$ Max Planck Institute for Extraterrestrial Physics, 85741 Garching, Germany.
      \\
      $^s$ Aarhus University, Interdisciplinary Nanoscience Center (iNANO), 8000 Aarhus C, Denmark.
   }%
}%
\endnotetext[1]{Present address:~Bruker AXS GmbH, Karlsruhe, Germany}%
\endnotetext[2]{Present address:~ARC Centre of Excellence for Coherent X-ray Science, School of Physics, The University of Melbourne, Australia}%
\endnotetext[3]{Present address:~Physikalisch-Technische Bundesanstalt, Bundesallee 100, 38116 Braunschweig, Germany}%
\endnotetext[4]{Present address:~Stanford PULSE Institute, SLAC National Accelerator Laboratory, 2575 Sand Hill Road, Menlo Park, California 94025, USA}%
\endnotetext[5]{Present address:~European X-ray Free Electron Laser (XFEL)GmbH, 22761 Hamburg, Germany}%

\clearpage

\section{Introduction}
\label{sec:intro}
The advent of X-ray Free-Electron Lasers (XFELs) opens up new and previously inaccessible research
directions in physical and chemical sciences. One of the major scopes is the utilization of XFEL
radiation in diffractive imaging experiments. Collecting single-shot x-ray diffraction patterns with the
ultrashort, currently down to a few femtoseconds, x-ray pulses of extremely high brilliance at an
XFEL allows the conventional damage limit in imaging of non-crystalline biological samples to be
circumvented.\cite{Neutze:Nature406:752} Experiments at the Linac Coherent Light Source (LCLS) confirmed
the feasibility of utilizing XFELs for femtosecond single-shot imaging of non-crystalline biological
specimens\cite{Seibert:Nature470:78} as well as for femtosecond nanocrystallography of
proteins.\cite{Chapman:Nature470:73}

These results provide important steps on the path towards the paramount goal of atomically
(picometer and femtoseconds) resolved diffractive imaging of structures and ultrafast structural
dynamics during chemical reactions of even single molecules. However, the path toward this goal,
often nicknamed as ``recording of a molecular movie'', is still long and many challenges have to be
overcome in order to achieve the required
spatio-temporalresolution.\cite{Chergui:CPC10:28,Barty:ARPC64:415} The usually proposed experimental
approach is to provide identical molecules, delivered in a liquid or gaseous stream to the focus of
an XFEL.\cite{Spence:PRL92:198102, Filsinger:PCCP13:2076} Since the high single-shot XFEL intensity
by far exceeds the damage threshold of single molecules, the molecules have to be replenished in
each shot. Single-molecule diffraction data has to be collected for many shots with the molecule at
many different orientations in order to fill up the three-dimensional diffraction volume. The
relative orientation of single-molecule diffraction patterns from distinct shots could be determined
computationally from the diffraction patterns themselves provided that the single-molecule
diffraction signal is well above noise.\cite{Loh:PRE80:026705, Fung:NatPhys5:64,
   Yefanov:JPB46:164013} However, one of the main issues in single-molecule x-ray diffraction
experiments is the weak scattering signal from single molecules, which, so far, is too weak to allow
for orientation classification solely from the diffraction pattern, even at the high intensities of
the novel XFELs. Therefore, diffraction data has to be recorded and averaged for many shots with the
molecule at the same, pre-imposed alignment and/or orientation \footnote{Alignment refers to fixing
   one or more molecular axes in space, while orientation refers to breaking of the corresponding
   up-down symmetry.} in space in order to obtain an interpretable diffraction pattern above noise.
Strong molecular alignment in the laboratory frame can be achieved, for instance, through adiabatic
laser alignment, while orientation requires additional dc electric fields.\cite{Larsen:PRL85:2470,
   Stapelfeldt:RMP75:543, Holmegaard:PRL102:023001, Nevo:PCCP11:9912}. Alignment and orientation can
be varied easily by controlling the the alignment laser polarization and, in case orientation is
utilized as well, the direction of the dc field. Utilizing ensembles of such aligned molecules
allows for averaging of many identical patterns, similar to recent experiments exploiting electron
diffraction from CF$_3$I~[\onlinecite{Hensley:PRL109:133202}] or photoelectron imaging of
1-ethynyl-4-fluorobenzene\cite{Boll:PRA88:061402, Rolles:FD171:inprep} and
dibromobenzene.\cite{Rolles:JPB:inprep}

An obstacle to this concept is that complex large molecules typically exist in various structural
isomers, \eg, conformers, which are often difficult to separate due to the small energy difference and
low barriers between them. However, to achieve atomic-resolution in diffractive imaging experiments
they have to be analyzed separately. We have proposed\cite{Filsinger:PCCP13:2076} to solve this by
spatially separating shapes,\cite{Helden:Science267:1483} sizes,\cite{Trippel:PRA86:033202} or
individual isomers\cite{Filsinger:PRL100:133003, Filsinger:ACIE48:6900, Kierspel:CPL591:130} of the
molecules before delivery to the interaction point of the experiment. These pre-selected ensembles
can be efficiently, one- and three-dimensionally, aligned or oriented in the laboratory
frame.\cite{Holmegaard:PRL102:023001, Filsinger:JCP131:064309, Trippel:MP111:1738}

Here, we give a detailed account of an x-ray diffraction experiment of ensembles of isolated
gas-phase molecules at the Linac Coherent Light Source (LCLS).\cite{Kuepper:PRL112:083002} Cold,
state-selected, and aligned ensembles of the prototypical molecule 2,5-diiodobenzonitrile
(\DIBNformula, DIBN) were irradiated with XFEL pulses with a photon energy of 2~keV
($\lambda=620$~pm) and x-ray diffraction data was recorded and analyzed. DIBN was utilized for this
proof-of-principle experiment because it contains two heavy atoms (iodine) and it can be
laser-aligned along an axis almost exactly coinciding with the iodine-iodine axis. Therefore, as the
two-center iodine-iodine interference dominates the scattering signal, the experiment resembles
Young's double slit on the atomic level. We achieved strong laser-alignment of the ensemble of DIBN
molecules which allowed for averaging of many patterns from these weakly scattering molecules.

The outline of this paper is as follows: In \autoref{sec:experimental} the experimental setup is
introduced. This includes details on the preparation of the molecular sample for the x-ray
diffraction experiment: we present measurements of the molecular beam deflection profiles and
two-dimensional ion-momentum distributions from which the molecular alignment of DIBN is quantified.
In addition, the process of data acquisition, background subtraction, and spatial single-photon
counting with the pnCCD photon detector\cite{Strueder:NIMA614:483, Hartmann:NSSCR:2590} is outlined
very briefly, while a comprehensive explanation of all the steps involved in the procedure of
conditioning and correcting the x-ray diffraction data is given in \ref{sec:appendix:daq}. The
theory behind the numerical simulations of x-ray diffraction intensities to be compared with the
experimental diffraction data is outlined in \autoref{sec:simulation}. In \autoref{sec:results}
the experimental results are presented and the manuscript concludes with a summary of the
experimental findings and an outlook on future experiments in \autoref{sec:outlook}.

\section{Experimental}
\label{sec:experimental}
\subsection{Experimental setup}
\label{sec:experimental:setup}
The experiment was performed at the Atomic, Molecular, and Optical Physics (AMO)
beamline\cite{Bozek:EPJST169:129, Bostedt:JPB46:164003} of LCLS,\cite{Emma:NatPhoton4:641}
using the CAMP (CFEL-ASG Multi-Purpose) experimental chamber.\cite{Strueder:NIMA614:483,Foucar:CPC183:2207}
The CAMP instrument was equipped with a state-of-the-art molecular beam setup providing gas-phase ensembles
of cold and quantum-state selected target molecules.\cite{Filsinger:JCP131:064309,Filsinger:PRL100:133003,
Filsinger:ACIE48:6900, Kierspel:CPL591:130,Trippel:PRA86:033202} For the x-ray diffraction
experiment, a photon energy of 2~keV ($\lambda=620$~pm) was used, which is the maximum photon energy
available at AMO. The 2~keV x-ray pulses were focussed by a  Kirkpatrick-Baez (KB) mirror system
into the CAMP
experimental chamber, which was attached to the High Field Physics (HFP) chamber at the AMO
beamline. The CAMP instrument contains multiple detectors to detect photons, electrons, and ions
simultaneously and it is described in detail elsewhere.\cite{Strueder:NIMA614:483}

\begin{figure}[t]
   \centering
   \includegraphics[width=\textwidth]{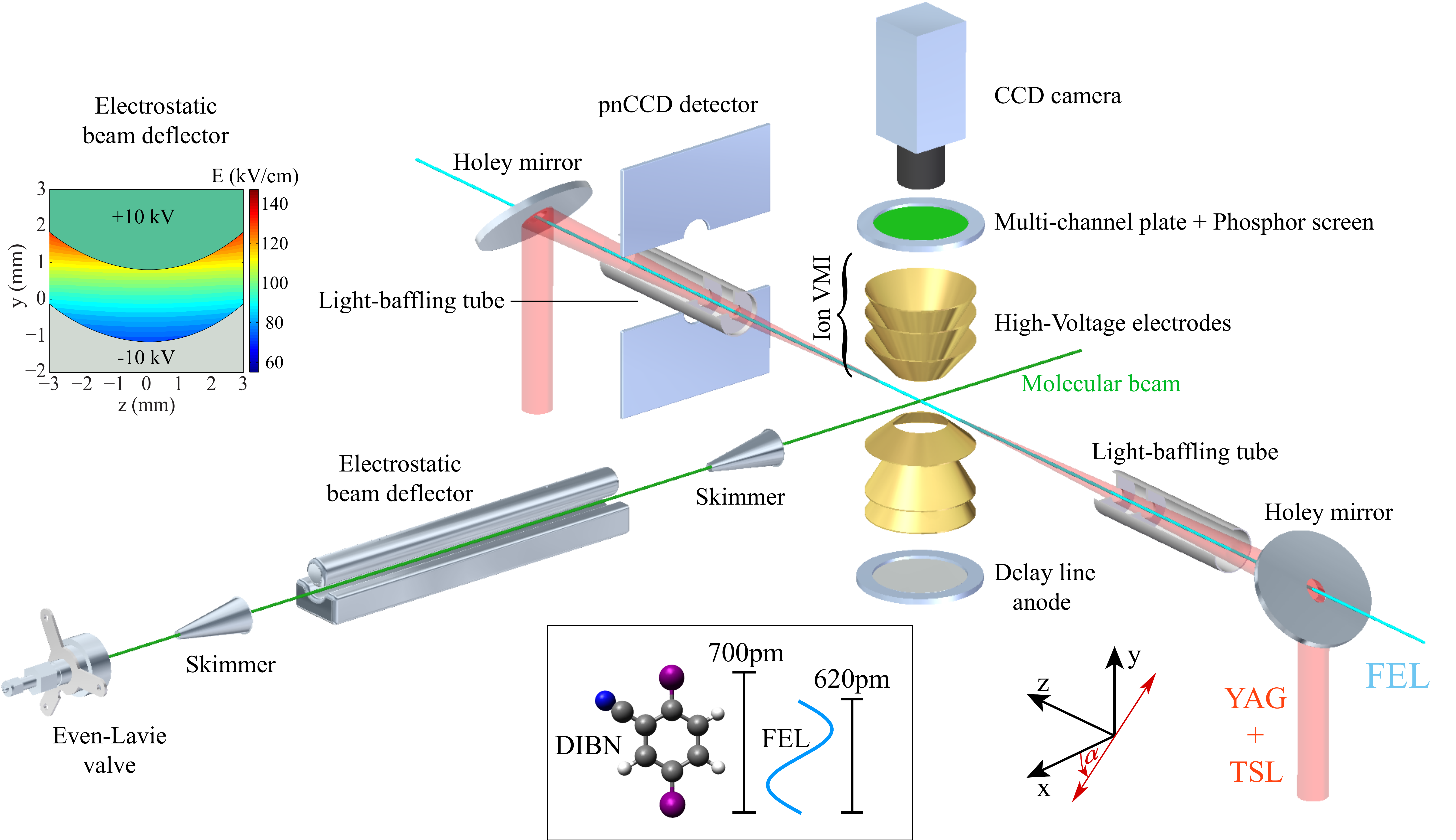}
   \caption{Schematic view of the experimental setup inside the CAMP experimental chamber. The
      molecular beam, created by supersonic expansion of DIBN and He from the Even-Lavie valve on
      the left, enters the deflector and quantum-state selected molecules are delivered to the
      interaction point. In the center of the velocity map imaging spectrometer (VMI) the molecular
      beam is crossed by the laser beams copropagating from right to left. The direct laser beams
      pass through a gap in the pnCCD photon detectors that are used to record the x-ray diffraction
      pattern. The upper pnCCD panel is further away from the beam axis than the bottom panel in
      order to cover a wider range of scattering angles. The inlet on the upper left shows a cross
      section of the electrostatic beam deflector along the propagation direction of the molecular
      beam. The inlet on the lower edge illustrates the two significant lengthscales of the x-ray
      diffraction experiment, namely the molecular structure of DIBN with the iodine-iodine distance
      and the x-ray wavelength. The molecular structure of DIBN was obtained from \emph{ab initio}
      calculations (GAMESS-US,\cite{Schmidt:JCC14:1347} MP2/6-311G**), which predict a value of
      700~pm for the iodine-iodine distance. Figure reproduced from ref. \onlinecite{Kuepper:PRL112:083002}.}
   \label{fig:setup}
\end{figure}
\autoref{fig:setup} shows a schematic view of the experimental setup inside CAMP. During the experiment,
a pulsed molecular beam was formed by a supersonic expansion of a mixture of a few mbar of DIBN and 50~bar
of helium (He) into vacuum through an Even-Lavie valve.\cite{Even:JCP112:8068} The target molecules were
cooled to low rotational temperatures of $\sim$1~K in the early stage of the expansion by collisions with
the He seed gas.\cite{Scholes:AtomMolBeam, Luria:JPCA115:7362} Traveling through the electrostatic
deflector, the molecules were dispersed along the vertical ($y$) axis according to their effective dipole
moment,
\ie, their quantum state. The deflector consists of two $24$~cm-long electrodes, a cylindrical rod
electrode at the top and a trough electrode at the bottom. The vertical distance between the two
electrodes in the horizontal center of the deflector is $2.3$~mm. By application of high static
electric potentials of $\pm10$~kV to the top and bottom electrodes, a strong inhomogeneous static
electric field was created with an electric field strength of 120~kV/cm and an electric field
gradient of $250$~kV/cm$^{2}$ in the center of the deflector as depicted in the inlet of
\autoref{fig:setup}. Quantum-state selection via the deflector is achieved due to the different
Stark effect of distinct quantum states (\emph{vide infra}). Furthermore, spatial separation of
polar DIBN and non-polar He seed gas in the deflector was utilized to reduce the scattering
background from the He in the x-ray diffraction experiment.

After passing through the deflector, the quantum-state dispersed molecular beam entered the detection
chamber where it was crossed by three pulsed laser beams: Pulses from a Nd:YAG laser (YAG, $12$ ns (FWHM),
$\lambda=1064$~nm, $E_I=200$~mJ, $\omega_0=63$~$\mu$m, $I_0\approx\!2.5\cdot10^{11}$~W/cm$^2$) were used
to align the ensemble of target molecules. The second laser, a Ti:Sapphire laser (TSL, $60$~fs (FWHM),
$800$~nm, $E_I = 400$~$\mu$J, $\omega_0=40$~$\mu$m, $I_0\approx\!2.5\cdot10^{14}$~W/cm$^2$) was used to
ionize DIBN in order to optimize the molecular beam and the alignment without the LCLS beam. X-ray pulses
from LCLS ($100$~fs, estimated from electron bunch length and
pulse duration measurements,\cite{Duesterer:njp13:093024} $\lambda=620$~pm, $E_I=4$~mJ,
$\omega=30$~$\mu$m, $I_0\approx2\cdot10^{15}$~W/cm$^2$) were used to probe the ensemble of aligned
DIBN. We deliberately worked out-of-focus of the x-ray beam at low fluence in order to mitigate
electronic\cite{Lorenz:PRE86:051911, Ziaja:NJP14:115015, Fratalocchi:PRL106:105504} and nuclear
damage processes.\cite{Erk:PRL110:053003} The x-ray photons diffracted from the ensemble were
collected by the pnCCD photon detector at a distance (\ie, camera length) of $71$~mm. $35$\% of the
generated $1.25\cdot10^{13}$ x-ray photons/pulse were estimated to be transported to the
experiment.\cite{Rudek:NatPhoton6:858} The two panels of the pnCCD detector were opened by a
significant amount in order to cover large scattering angles, \ie, the top pnCCD panel was moved by
44~mm (covering scattering angles of $31\degree\leq\twotheta\leq 50\degree$) and the bottom panel to
a distance of 17~mm ($13\degree\leq\twotheta\leq 38\degree$) from the $z$-axis. All three laser
beams were co-propagating, overlapped using dichroic (1064~nm and 800~nm) and holey (x-ray and
infrared beams) mirrors. After intersecting the sample the lasers finally left the setup through a
gap between the two panels of the pnCCD camera and another holey mirror in the back of the CAMP chamber
to separate the laser beams again. Straylight from the optical lasers was reduced using a set of apertures
mounted in a small tube directly in front of the interaction zone (named ``light baffling tube'' in
\autoref{fig:setup}). A similar light baffling tube was mounted downstream the interaction zone,
reaching between the two pnCCD panels and containing a similar set of apertures in order to suppress
straylight from optical or x-ray photons impinging from the back of the CAMP chamber onto the back
of the pnCCD panels. In addition, the front side (\ie, the side facing the interaction zone) of each pnCCD
panel was covered using aluminum-coated filters in order to further suppress straylight from the
optical lasers.

The CAMP chamber was equipped with a dual velocity-map-imaging (VMI) spectrometer in order to measure
two-dimensional ion momentum distributions in the x-z plane, resulting from Coulomb explosion due to
absorption of one or a few x-ray photons (or optical photons in case the TSL was utilized to probe the
molecular alignment).\cite{Strueder:NIMA614:483} Operation of the VMI spectrometer as an ion time-of-flight
(TOF) spectrometer in quasi-Wiley-McLaren configuration\cite{Wiley1955} allowed for mass selective detection
of individual ionic fragments.

The x-ray diffraction experiment was performed with LCLS running at a repetition rate of 60~Hz while
the YAG was running at 30~Hz. Hence, a dataset contains shots of aligned and randomly oriented molecules
in an alternating manner. All diffraction measurements were conducted in the deflected part of the molecular
beam, \ie, at (nearly) optimal molecular alignment (\emph{vide infra}). In the following, experimental
results concerning preparation of the molecular ensemble for the x-ray diffraction experiment are presented,
namely quantum-state selection by deflection and laser-alignment.

\subsection{Quantum-state selection and laser alignment}
\label{sec:experimental:deflection_and_alignment}
The benefit of quantum-state selection prior to laser alignment for cold ensembles of asymmetric top
molecules\cite{Holmegaard:PRL102:023001} was exploited in our experiment in order to obtain strong
alignment of the molecular sample for the x-ray diffraction experiment. For a large asymmetric top
molecule such as DIBN, all populated rotational states in the molecular beam are so-called
high-field-seeking (hfs) states. Molecules in these states are deflected towards increasing electric
field strength, \ie, upwards along the $y$-axis.\cite{Filsinger:JCP131:064309, Chang:CPC185:339} The
lowest states typically exhibit the largest Stark energy shift and, thus, the strongest deflection.
Quantum-state selection is very beneficial for laser-alignment: As the lowest-lying states
experience a stronger angular confinement in the electric field of a linearly polarized alignment
laser, selection of the lowest-lying states prior to alignment significantly improves the degree of
alignment.\cite{Holmegaard:PRL102:023001, Filsinger:JCP131:064309, Trippel:MP111:1738}

When the linearly polarized YAG was included, DIBN molecules aligned along their most-polarizable
axis, which is nearly coincident with the iodine-iodine (I--I) axis. Utilizing Coulomb explosion
imaging of aligned DIBN, induced by either the TSL or the FEL, strong alignment of DIBN ensembles
was confirmed by two-dimensional momentum distributions of \Iplus\ ions (which recoil along the
iodine-iodine axis) recorded with the velocity-map imaging (VMI) spectrometer.
\begin{figure}
   \centering
   \includegraphics[width=0.70\textwidth]{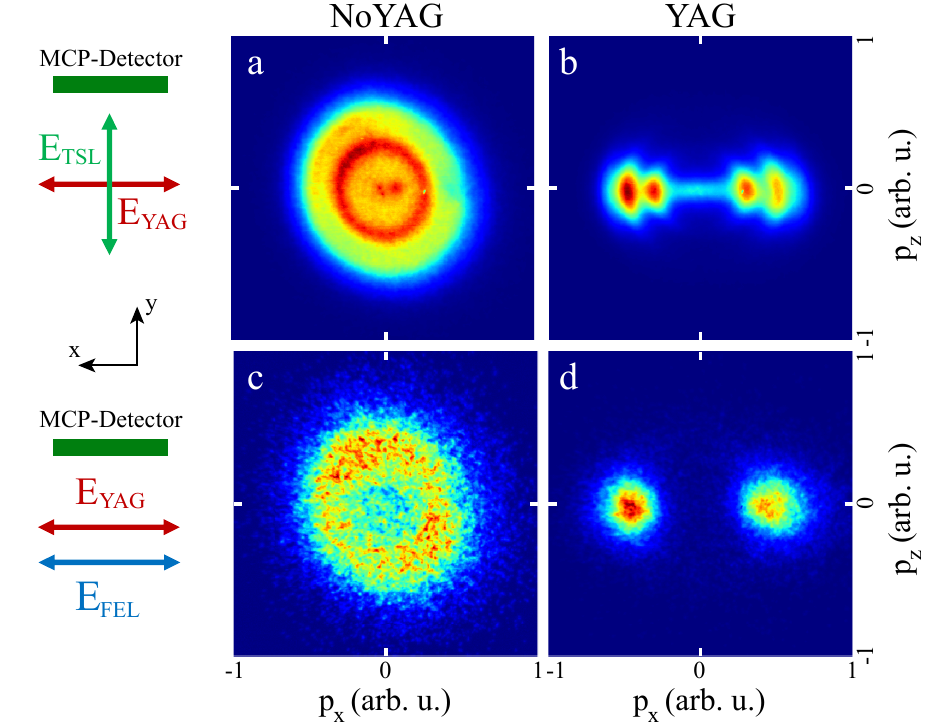}
   \caption{\Iplus\ ion momentum distributions recorded with the ion-VMI and MCP detector when
      the TSL (a,b) or the LCLS (c, d) was used to ionize and Coulomb explode the molecules. In (a, c)
      cylindrically symmetric distributions from isotropic ensembles are observed (the images are
      slightly distorted due to varying detector efficiencies). In (b, d) the horizontal alignment
      of the molecules, induced by the YAG, is clearly visible. In all measurements the YAG and the
      LCLS are linearly polarized along the $x$-axis, \ie, parallel to the detector plane, and the TSL is
      linearly polarized along the $y$-axis, \ie, perpendicular to the detector plane. Figure reproduced
      from ref. \onlinecite{Kuepper:PRL112:083002}.}
   \label{fig:alignment}%
\end{figure}
\autoref{fig:alignment} shows corresponding \Iplus\ momentum distributions, recorded with (YAG) and
without (NoYAG) the YAG alignment laser. In the NoYAG case, the \Iplus\ images are circularly
symmetric corresponding to an ensemble of isotropically-distributed molecules. The circularly
symmetric image \autoref{fig:alignment}\,c, obtained following ionization with the horizontally
polarized FEL also demonstrated that the interaction of the far-off resonant radiation with the
molecule was independent of the angle between the molecular axis and the x-ray polarization
direction: The x rays were a practically unbiased ideal probe of spatial orientation of molecules.
Including the YAG laser, \Iplus\ ions were strongly confined along the polarization axis of the YAG.
The two distinct pairs of peaks in the TSL case correspond to two distinct ionization channels
yielding \Iplus\ ions from doubly and triply ionized molecules.\cite{Larsen:JCP111:7774} The degree
of alignment is quantified by calculating \cost, where \thetadd\ is the angle with respect to the
laser polarization axis in the projected, two-dimensional \Iplus\ momentum distributions.

\begin{figure}[t]
   \centering
   \includegraphics[width=0.60\textwidth]{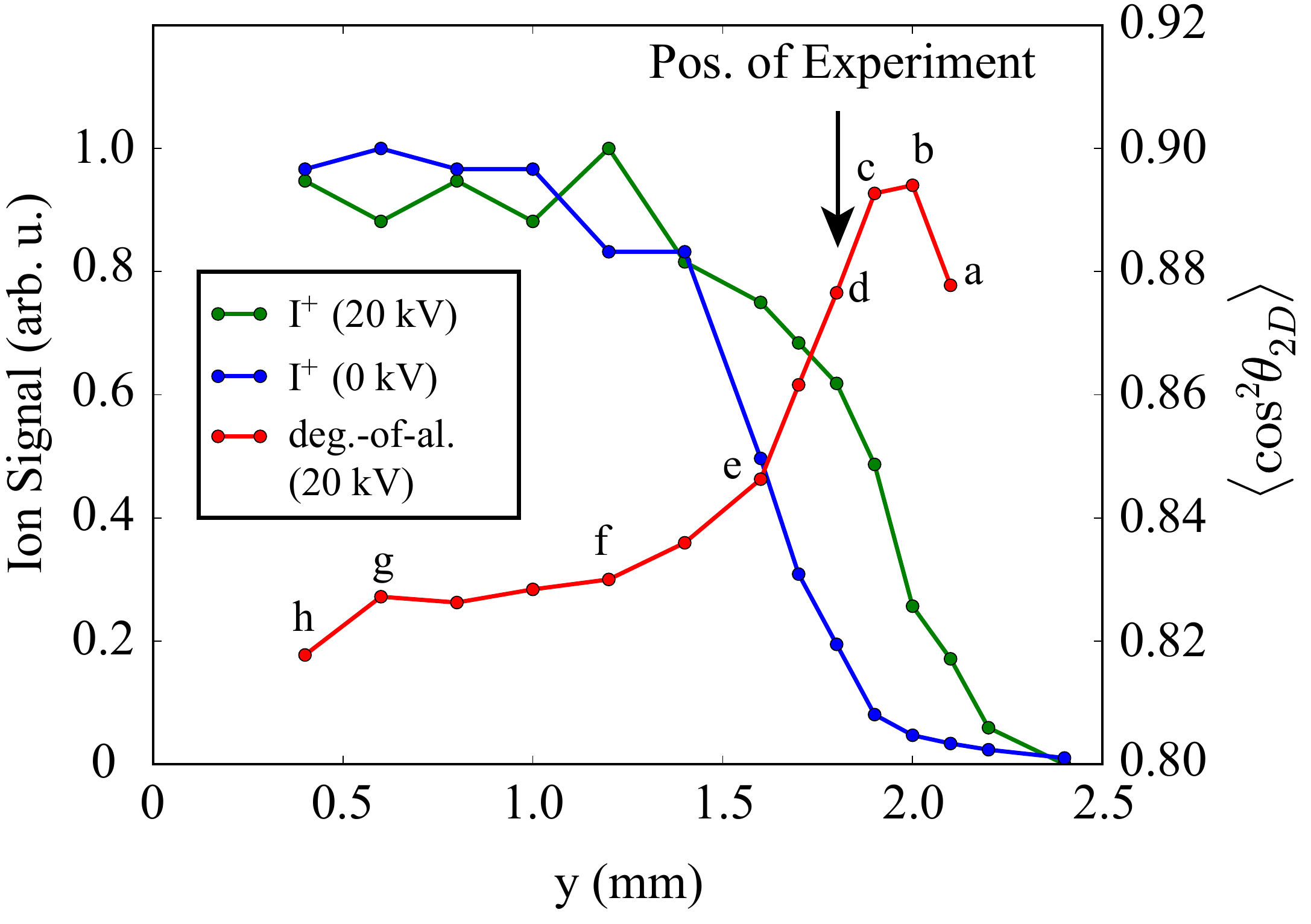}
   \caption{The molecular beam density profiles, obtained by recording the \Iplus\ signal
            (see left vertical axis) at different positions along the $y$-axis in the molecular
            beam for the undeflected (blue) and deflected (green) molecular beam. The different
            degree of alignment of DIBN in terms of \cost~(right vertical axis) at different
            positions in the deflected molecular beam illustrates the dispersion of quantum
            states (red). Considering the best compromise between degree of alignment and
            sufficient molecular beam density of target molecules, the x-ray diffraction experiment
            was performed at $y=1.8$~mm ($\cost=0.877$), not at the position where the highest
            degree of alignment was observed, \ie, $\cost=0.894$ at $y=2$~mm.}
   \label{fig:deflection_alignment}
\end{figure}
The deflector was utilized to improve the degree of alignment by quantum-state selection of the
lowest states. \autoref{fig:deflection_alignment} shows molecular beam density profiles obtained by
measuring the \Iplus\ signal probed at distinct positions along the $y$-axis when the deflector was
off (blue) or on (green, $20$~kV). Both graphs were normalized to the peak intensity. Only the upper
part of the molecular beam was probed. The different deflection of distinct quantum states in the
molecular beam leads to a shift of the beam profile as is shown in \autoref{fig:deflection_alignment}.
The corresponding dispersion of quantum states can be illustrated by recording \Iplus\ momentum
distributions at distinct positions: as expected, the degree of
alignment is significantly enhanced in the deflected part of the molecular beam. The resulting
\cost\ values are depicted by the red graph of \autoref{fig:deflection_alignment} and the enhanced
alignment in the deflected part of the molecular beam is obvious. The strongest alignment, quantified
by $\cost=0.894$, was obtained at $y=2$~mm. However, to utilize a higher beam density, the x-ray
diffraction experiment was performed at $y=1.8$~mm. At this position the degree of alignment was only
slightly smaller ($\cost=0.877$), but the molecular beam density was still 60~\% of the undeflected beam
density, whereas it was only 20~\% at $y=2$~mm.

During the x-ray diffraction experiment, the YAG polarization was rotated to $\alpha=-60\degree$
with respect to the horizontal axis. The alignment was probed repeatedly over the course of the
x-ray diffraction measurement period of $\sim8$~h and the average degree of alignment was
$\cost=0.84$, mainly due to variations of the overlap of the YAG and FEL pulses. The degree of
alignment is in agreement with measurements of adiabatic alignment of quantum-state selected
ensembles of similar molecules\cite{Holmegaard:PRL102:023001,Trippel:MP111:1738} and matches
requirements for diffraction experiments on aligned molecules.\cite{Filsinger:PCCP13:2076,
   Hensley:PRL109:133202}

\subsection{X-ray diffraction data acquisition}
\label{sec:experimental:daq}
A comprehensive description of the data conditioning procedure is given in \ref{sec:appendix:daq}.
In summary, single shot x-ray diffraction data was recorded by the pnCCD detectors and saved to
file. Several sources of background signals (offset, gain, experimental background from the
YAG alignment laser, etc.) and detector artifacts (``hot-pixels'', etc.) were subtracted from the
data by utilizing the CFEL-ASG Software Suite (CASS).\cite{Foucar:CPC183:2207} Eventually, single
x-ray photon hits were extracted by application of a $3\sigma$-threshold to these ``clean'' single-shot
pnCCD data frames. This procedure yields 0.2 x-ray photons per shot (\ie, on average only one scattered
x-ray photon in five shots) that are scattered to the pnCCD detector. These photons are placed in a
histogram which represents the molecular diffraction pattern obtained from aligned (labelled ``YAG'')
and isotropically distributed (``NoYAG'') ensembles of DIBN molecules.

\section{Simulation of x-ray diffraction intensities}
\label{sec:simulation}
Diffraction intensities from ensembles of aligned and not-aligned DIBN molecules and the He seed gas
were simulated for comparison with the experimental data. Unless stated otherwise, the underlying
theory is either explicitly given by the book of Als-Nielsen \&
McMorrow\cite{Als-Nielsen:ModernXrayPhysics} or was derived from there.\cite{Stern:thesis:2013}

X-ray scattering off ensembles of isolated molecules is very weak and hence the kinematical
approximation (first Born approximation) is assumed to be valid, meaning that multiple scattering of
a single photon is highly unlikely and can be neglected. For all calculations, the interaction point
is regarded as the origin of the coordinate system. Then, the number of x-ray photons
$I_{\text{sc}}$ that are scattered from a single molecule to a certain pixel at position
$\mathbf{R}$ can be calculated as
\begin{equation}
   I_{\text{sc}} = \left[r_{0} \cdot F_{\text{mol}}(\mathbf{q}) \cdot
      e^{i\mathbf{kR}}\; \right]^2 \cdot \Delta \Omega \cdot P \cdot \frac{I_{0}}{A_0}
   \label{eq:intensity}
\end{equation}
where $r_{0}$ is the Thomson scattering length of the electron which is given by
\begin{equation}
   r_{0} = \frac{e^{2}} {4\pi \epsilon_{0} m c^{2}} \;\; = \;\; 2.82\cdot 10 ^{-5} \; \textrm{\AA}
   \label{eq:ThomsonElectron}
\end{equation}
Utilizing conventional notation, $\mathbf{q}=\mathbf{k}-\mathbf{k}'$ is the scattering vector with
$\mathbf{k}$ and $\mathbf{k}'$ being the wavevectors of the incident and scattered waves,
respectively. $F_{\text{mol}}(\mathbf{q})$ is the molecular scattering factor (see below).
$\Delta\Omega$ is the solid angle a certain pixel subtends to the incident XFEL beam, and $P$ is the
polarization factor depending on the x-ray source. Since LCLS is linearly polarized (along the
$x$-axis), $P$ takes the following form:
$P(\mathbf{k}')=1-|\hat{\mathbf{u}}\cdot\hat{\mathbf{k}}'|^2$ with the unit vector
$\hat{\mathbf{u}}$ pointing along the $x$-axis.\cite{Kirian:OE18:5713} Finally, the number of
incident photons is given by $I_{0}$ and the cross-sectional area of the incident x-ray beam is
represented by $A_0$.

The scattering factor of a molecule $F_{\text{mol}}(\mathbf{q})$ is modeled as the sum of the atomic
scattering factors $f_{j}(\mathbf{q})$ of the constituent $j$ atoms (located at the positions
$\mathbf{r}_j$ within the molecule) times the phase factor $e^{i\mathbf{qr}_j}$, hence
\begin{equation}
   F_{\text{mol}}(\mathbf{q}) = \sum\limits_{j} f_{j}(\mathbf{q}) \;\; e^{i\mathbf{qr}_j}
   \label{eq:Fmol}
\end{equation}

A model of the atomic scattering factors $f_{j}$ has been given by Waasmaier \&
Kirfel\cite{Waasmaier:ActaCrystA51:416} by modelling atomic scattering factors in dependence of the
scattering momentum transfer $s=\sin\Theta/\lambda$ as the sum of five gaussian functions and a
constant.
\begin{equation}
   f(s) = \sum\limits_{i=1}^{5} a_{i}\;e^{-b_{i}\:s^2}\;+\; const.
   \label{eq:Waasmaier}
\end{equation}
For the calculations presented here, the atomic scattering factors were modified by dispersion
corrections given by Henke \etal,\cite{Henke:ADNDT54:181} thereby accounting for the dependence of
the scattering strength from the photon energy.

\eqref{eq:intensity} was used to calculate the diffraction pattern for a perfectly aligned molecule.
However, the experimental diffraction pattern of an ensemble of DIBN molecules with a finite (\ie,
non-perfect) degree of alignment is the incoherent superposition of single-molecule diffraction
patterns at slightly different orientations with respect to the (linear) laser polarisation of the
YAG. The relative weight of different orientations are described by an alignment-angular
distribution function giving the relative population $n(\theta)$ where $\theta$ is the angle with
respect to the YAG polarisation axis. The following approximation for strong alignment was applied
in our model:\cite{Friedrich:PRL74:4623}
\begin{equation}
   n(\theta) = \exp{\left(-\frac{\sin^2\theta}{2\sigma^2}\right)}
   \label{eq:AlignmentAngularDistribution}
\end{equation}

\begin{figure}[t]
   \centering
   \includegraphics[width=1.00\textwidth]{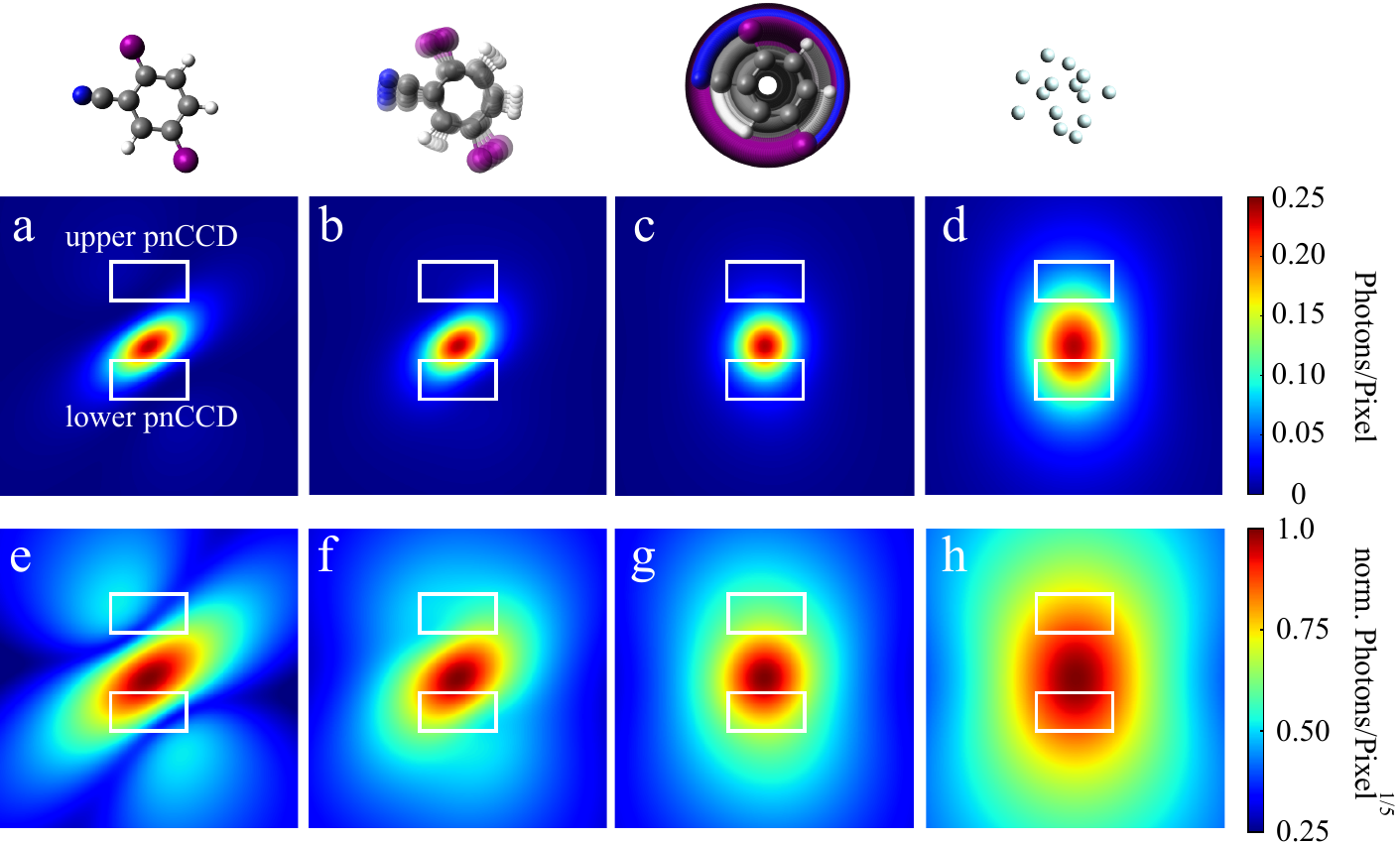}
   \caption{Simulated scattering intensities for different degrees of alignment. a--d correspond to
      DIBN aligned with $\cost=0.99$ (a), $0.84$ (b), $0.5$ (isotropic,c). The signal for 5\,580 He
      atoms (d) is the same as the DIBN signal at $\mathbf{q} = 0$. To illustrate interference features
      (\ie, the weak first order diffraction maxima), the second row (e--h) shows the fifth root of
      the normalized intensities of the first row.}
   \label{fig:simulation}%
\end{figure}

In practice, the blurred single-molecule diffraction pattern was obtained by averaging of
single-molecule diffraction patterns calculated for 1000 distinct orientations of DIBN, weighted by
\eqref{eq:AlignmentAngularDistribution}. Then, this pattern is multiplied by the number of
molecules $N$ in the interaction volume $V_0$ in order to obtain the diffraction pattern of $N$
molecules. However, as long as the x-ray beam is smaller than the molecular beam, the absolute
number doesn't have to be known but rather the number density $M$ of molecules: The number of
molecules $N$ can be written as $N=M\cdot{}V_0=M\cdot{}A_0\cdot{}l$ where the interaction volume is
approximated as a cylindrical volume of lenght $l$ in $z$-direction, and, in our case, $l$ is the width
of the molecular beam which is $\approx\!4$~mm (determined by the last skimmer). Therefore, once
\eqref{eq:intensity} was multiplied by $N$, the factor $N/A_0$ in \eqref{eq:intensity} could be
replaced by $M\cdot{}l$.

\autoref{fig:simulation} shows simulated diffraction patterns, \ie, the number of scattered photons
on a plane detector at a camera length of 71~mm, for different degrees of alignment for 565\,000
shots ($4.375\cdot10^{12}$ photons/shot), and a molecular beam density of
$\M=1.2\cdot10^{8}\text{~cm}^{-3}$. The molecules were aligned at $\alpha=-60\degree$ with respect
to the horizontal plane. White rectangles mark the position of the pnCCDs in the experiment. Images
a--c correspond to DIBN aligned with $\cost=0.99$ (a), 0.83 (b), 0.5 (isotropic, c). The diffraction
signal from 5\,580 He atoms (d) at $\mathbf{q}=0$ is equal to the diffraction signal from a single
DIBN molecule at $\mathbf{q}=0$. We do not exactly know the ratio of He atoms per DIBN molecule in
our molecular beam, but it is in the 10$^4$-10$^5$ range. In order to illustrate interference
features of the weak first order diffraction maxima, a different colorscale has been applied,
enhancing the first-order iodine-iodine diffraction maxima: therefore, the second row (e--h) shows
the fifth root of the normalized intensities. The main interference feature,
originating in the interference of the two iodine atoms, is clearly visible for nearly perfect alignment,
\ie, in \autoref{fig:simulation} (e) while non-perfect alignment (f) significantly washes out the
interference features at high angles.

\section{Results and discussion}
\label{sec:results}
Diffraction patterns \INoYAG and \IYAG were constructed independently for isotropic (NoYAG) and
aligned (YAG) samples, respectively, by summing all photon hits in the energy range around 2~keV,
corresponding to 1500--3200~ADU (analog-to-digital unit, see \ref{sec:appendix:daq}) into a
two-dimensional histogram. The resulting images are shown in \autoref{fig:Finished_Pixels1-2}~c
and~d in \ref{sec:appendix:daq}. In addition to the diffraction signal from aligned DIBN, the
\INoYAG- and \IYAG-data\ contain
experimental background such as the isotropic atomic scattering from all individual atoms of DIBN,
scattering from the helium seed gas, scattering from residual gas in the chamber, and scattering at
apertures in the laser beam path. Since the scattering background from all these sources is the same
under NoYAG and YAG conditions, it cancels out when calculating \IYAGmNoYAG.

\begin{figure}[t]
    \centering
    \includegraphics[width=0.80\textwidth]{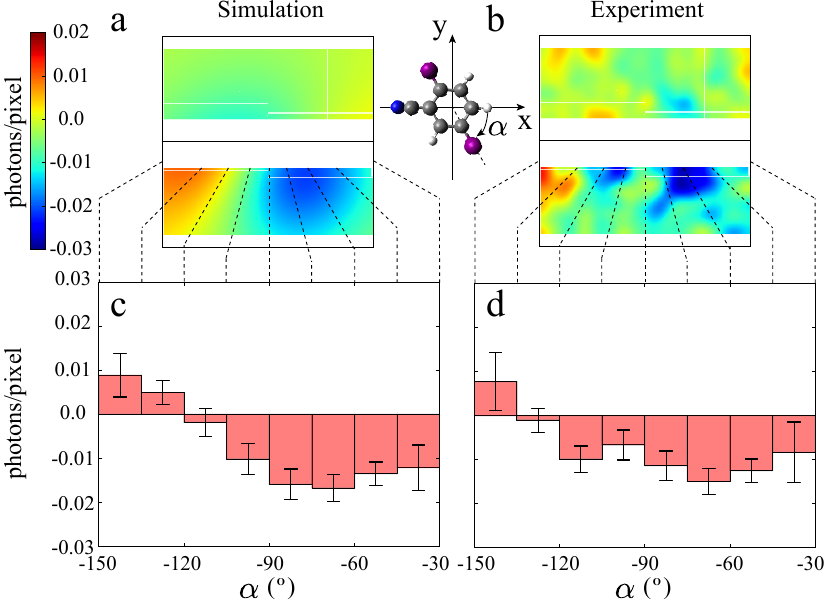}
    \caption{Diffraction-difference \IYAGmNoYAG\ of x-ray scattering in simulated (a) and
       experimental (b) x-ray-diffraction patterns. Histograms of the corresponding angular
       distributions on the bottom pnCCD (c,~d) illustrate the angular anisotropy of the diffraction
       signal. Error bars correspond to $1\sigma$ statistical errors from Poisson noise. The
       molecular beam density in (a) is $\M=0.8\cdot10^8$~cm$^{-3}$. Figure reproduced from
       ref. \onlinecite{Kuepper:PRL112:083002}.}
    \label{fig:data}
\end{figure}
\autoref{fig:data} shows the diffraction-difference pattern \IYAGmNoYAG\ for (a) simulated and (b)
experimentally recorded x-ray diffraction data. The \INoYAG\ data has been scaled to match the
number of shots of the \IYAG\ data. The difference is almost entirely due to the iodine-iodine
interference which dominates the anisotropic part of the scattering signal. The most notable
diffraction features are the zeroth-order maximum and the first-order minimum appearing on the
bottom pnCCD panel (\ie, at low resolution). The anisotropy of the diffraction signal of aligned
DIBN is illustrated by the angular anisotropy with respect to the alignment angle $\alpha$ as shown
in \autoref{fig:data}~c,~d. This anisotropy is well beyond statistical uncertainties, thereby
demonstrating x-ray diffraction signal from the aligned ensemble of isolated DIBN molecules.

Utilizing the iodine-iodine interference of the \IYAGmNoYAG pattern, it was investigated whether the
iodine-iodine distance could be estimated from the diffraction data. From \emph{ab initio}
calculations (GAMESS-US,\cite{Schmidt:JCC14:1347} MP2/6-311G**), a value of 700~pm was predicted for
the iodine-iodine distance. Taking into account the wavelength of 620~pm it is clear that the
interference features extend to high scattering angles \twotheta, \eg, the first scattering maximum
from the iodine-iodine interference appears at $\twotheta=51\degree$ which was not covered by the
detector; the outer corner of the top pnCCD panel corresponds to $\twotheta=50\degree$, \ie, the
resolution is low. For this reason, direct methods such as phase-retrieval from the diffraction
pattern alone were not applied. Instead, the data was compared to models of different iodine-iodine
distances and the best fit of a particular model to the data was estimated as will be explained in
the following.

\begin{figure}[t]
   \centering
   \includegraphics[width=1.00\textwidth]{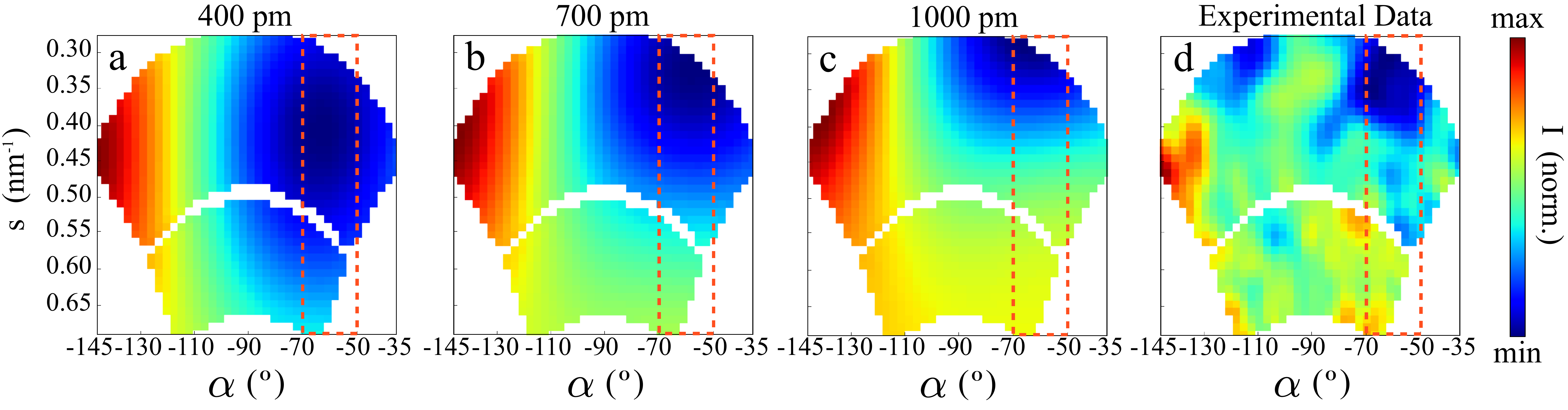}
   \caption{Diffraction difference \IYAG-\INoYAG\ in the $(s,\alpha)$-representation for simulated
      and experimental data. The simulated data shows the diffraction-difference \IYAG-\INoYAG\ for
      I--I distances of (a) 400~pm, (b) the theoretically expected I--I distance of 700~pm, and (c)
      1000~pm. The experimental data is shown in (d). The dashed red frames mark the azimuthal range
      $\alpha\in[-70\degree,-50\degree]$ along which the $I(s)$ graph is obtained.}
   \label{fig:YAG-NoYAG_s_phi_norm}
\end{figure}
\autoref{fig:YAG-NoYAG_s_phi_norm} shows the diffraction-difference \IYAG-\INoYAG\ in a different
representation. The $(x,y)$-coordinates were transformed to $(s,\alpha)$-coordinates, where
$s=\sin{\Theta}/\lambda$ is the scattering vector and $\alpha$ is the azimuthal angle. Due to the
twofold symmetry of the diffraction pattern for rotations about the z-axis, the upper pnCCD was
rotated by 180$\degree$ and ``connected'' to the bottom edge of the lower pnCCD, thereby extending
the range of $s$-values. Due to the masking of pnCCD regions during the generation of photon hit
lists (see \ref{sec:appendix:daq}) the active regions of the two pnCCD panels do not overlap.

\begin{figure}[b]
   \centering%
   \includegraphics[width=0.65\textwidth]{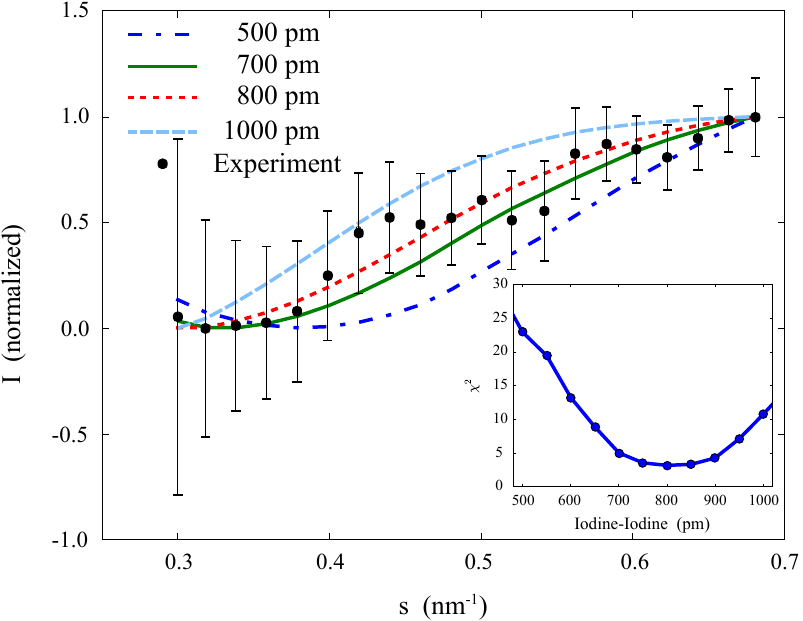}
   \caption{Comparison of experimentally obtained intensity profiles $I(s)$ along the alignment
      direction of the diffraction-difference pattern \IYAGmNoYAG with simulated profiles. The
      experimentally obtained $I(s)$ is best fitted (in terms of a $\chi^2$ test) with the model for
      an iodine-iodine distance of 800~pm. In the inset the test-statistic $\chi^2$ is shown in
      dependence of the iodine-iodine distance. Figure reproduced from ref. \onlinecite{Kuepper:PRL112:083002}.}
   \label{fig:analysis}%
\end{figure}
Varying the iodine-iodine distance \liod\ mainly results in squeezing/stretching of the diffraction
minima/maxima in the diffraction pattern. This is most pronounced along the alignment direction
$\alpha=-60\degree$ for the first diffraction minimum in our data. The intensity profile $I(s)$ of
the \IYAGmNoYAG\ data along with simulated $I(s)$ profiles for varying iodine-iodine distances is
shown as a function of the scattering vector $s$ in \autoref{fig:analysis}, averaged over
$-70\degree\le\alpha\le-50\degree$. Each graph is
normalized to be independent of the exact molecular beam density \M\ of DIBN molecules, which merely
changes the contrast, \ie, the depth of the minimum.

The agreement of the experimental data with a particular model is estimated in terms of a
\chisqr-test.\cite{Nakamura:JPG37:075021:33} The best fit to the data, corresponding to the minimum
\chisqr-value, is obtained for an iodine-iodine distance of 800~pm, see \autoref{fig:analysis}. Due
to the low resolution at the current experimental parameters, the fitting is not very accurate. Thus,
in future experiments the use of shorter wavelengths will be crucial for an accurate determination
of structural features with real atomic resolution. At LCLS, the shortest wavelength currently
available is $\lambda\approx130$~pm (photon energy $\sim\!9.5$~keV), while the European XFEL will be
able to provide radiation at wavelengths down to $\lambda\approx50$~pm (photon energy $>24$~keV)
from its start of operation in the near future.\cite{Altarelli:XFEL-TDR:2006} In addition,
the high repetition rate of 27\,000~Hz at European XFEL allows recordance of such diffraction
patterns with better statistics in even shorter amounts of time than is currently possible.
\\\\
Deviations from the equilibrium geometry could be explained by radiation damage effects, \ie,
nuclear and/or electronic damage due to the intense XFEL radiation. However, we estimate that
radiation damage effects could not be observed in our diffraction data. First, in contrast to
previous experiments explicitely investigating the radiation damage induced by strongly focused XFEL
beams,\cite{Rudek:NatPhoton6:858, Young:Nature466:56,Erk:PRL110:053003} we deliberately worked out
of focus (\ie, at $\omega = 30$~$\mu$m), thereby avoiding significant electronic damage effects.
Secondly, the wavelength of 620~pm and the range of recorded $s$-values is insufficient to resolve
nuclear motion during the 100~fs x-ray pulses. The reasoning is given in the following.

\begin{figure}[t]
   \centering%
   \includegraphics[width=0.65\textwidth]{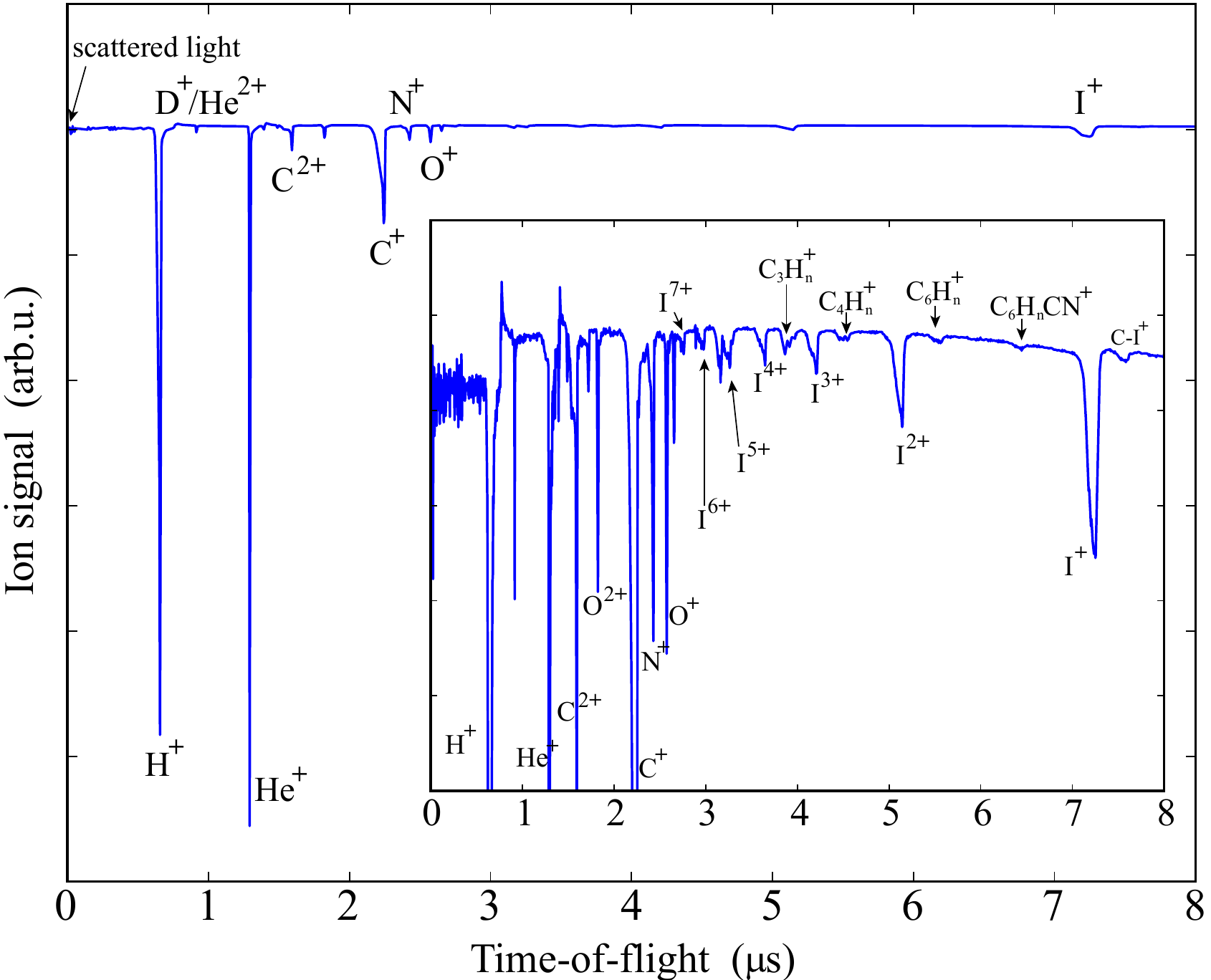}
   \caption{Time-of-flight spectrum, obtained by probing the molecular beam
   with the FEL. The inlet is a zoom into the vertical axis in order to show the
   various iodine ions.}
   \label{fig:tof}%
\end{figure}
\autoref{fig:tof} shows a time-of-flight spectrum, obtained by probing the molecular beam with the
FEL. Iodine ions with increasing charge ($\Iplus\ldots\Iplusseven$) appear in the spectrum with
decreasing intensity. In particular, singly-charged iodine \Iplus\ is most abundant while
fragments with charges higher than \Iplusseven are virtually absent in the spectrum. When DIBN is
ionized by 2~keV photons, predominantly the M-shell of iodine is accessed and the total
photo-ionization cross-section of iodine of $\sigma_{\text{abs}}=41.92~\text{pm}^2$ (0.4192~Mbarn)
is dominated by the cross-section of the $3p$ and $3d$ subshells. Considering the final charge
states reached via Auger decay upon photoabsorption of a 2~keV photon in the $3p$ and $3d$ subshells
of iodine, an Auger decay similar to xenon is expected, since the electronic decay processes do not
strongly depend on the atomic number. For xenon, multiply charged $\text{Xe}^{+n}$ ions are obtained
from such a photoionization event,\citealp{Kochur:JPB27:1709} \eg, an initial $3d$ vacancy in xenon
yields Xe$^{+4}$ as the most probable final charge state, while for a $3p$ vacancy, the charge-state
distribution is shifted upwards and peaks around Xe$^{+7}$. The most-probable final charge state
has, in both cases, a probability of $\approx\!50$~\%. Thus, by assuming similar ionization pathways
for xenon and iodine, the absorption of a single 2~keV photon by DIBN is likely to result in a
charge state distribution of DIBN peaking at DIBN$^{+4}$ or higher charges. Hence, iodine charge
states of $\Iplusone$ to $\Iplusseven$ could be entirely due to absorption of only a single photon.
We conclude that typically one photon is absorbed per molecule. In the following, absorption of two
or more photons is neglected.

For the moderate fluence conditions in our experiment, the probability $p_\text{abs}$ for single-photon
absorption of DIBN can be calculated based on the photoabsorption cross section of atomic iodine
$\sigma_{\text{abs}}=41.92~\text{pm}^2$ (0.4192~Mbarn).\cite{Berger:XCOM1.5:2010} Taking into
account the number of photons $N_{\text{photons}}=4.375\cdot10^{12}$ and the interaction area
$A_0=7.068\cdot10^{-10}~\text{m}^2$ ($706.8~\mu\text{m}^2$), the probability for photoabsorption of
a 2~keV photon by a single iodine atom is $p_\text{abs}=0.25$, hence the probability for a DIBN
molecule (\ie, two iodine atoms) is 0.5, \ie, half of the DIBN molecules absorb an x-ray photon, and,
eventually, become multiply ionized by Auger relaxation and fragment due to Coulomb explosion.

We estimate the influence of scattering from fragmenting DIBN on the diffraction pattern in terms of
a simple mechanical model concerning only nuclear damage, \ie, motion of ionic fragments happening
during the 100~fs (FWHM) x-ray pulse due to Coulomb explosion. The effective spatial distribution of
the two main scattering centers, \ie, the two iodine atoms, seen by the entire FEL during a single
shot is estimated, taking into account the gradual ionization during the course of the FEL pulse,
the total amount of ionization, and the velocity distribution obtained from the measured momentum
distributions of \Iplus\ ions.

\begin{figure}[t]
    \centering
    \includegraphics[width=0.8\linewidth]{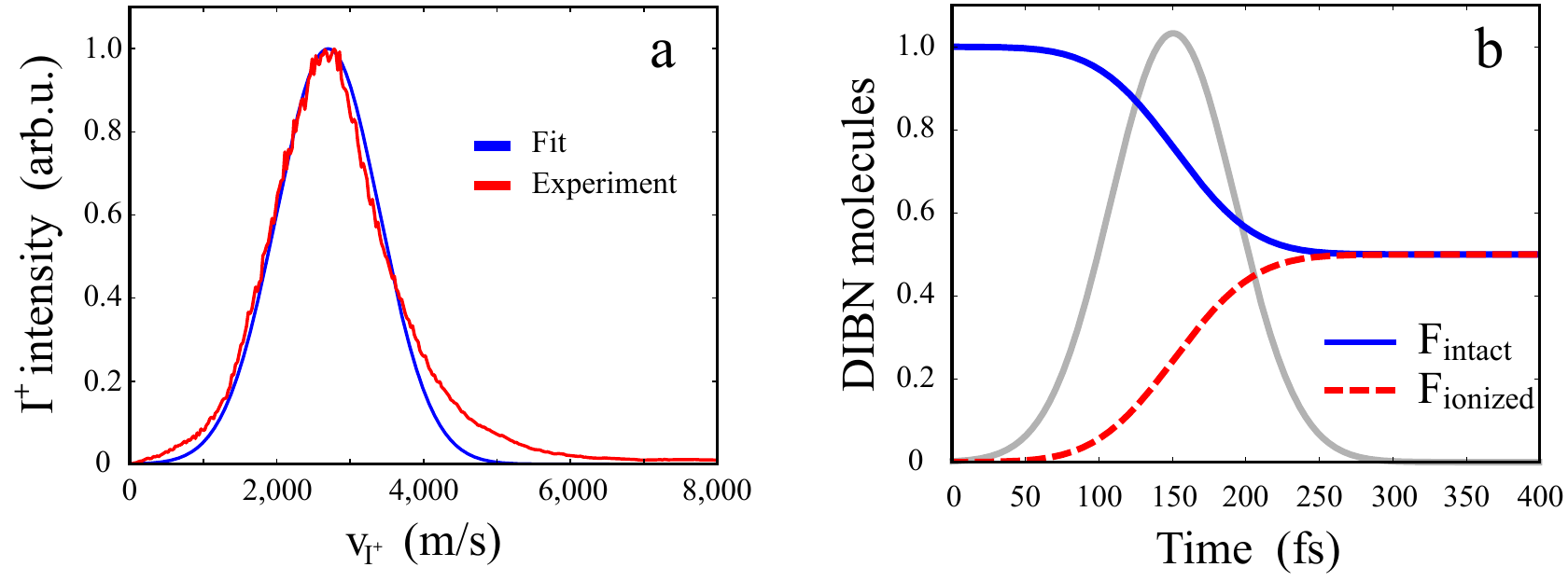}
    \caption{(a) One-dimensional velocity distribution of \Iplus\ ions, estimated from the momentum
       distributions as shown in \autoref{fig:alignment}. (b) The fractions of intact and ionized
       DIBN as a function of time for a 100~fs (FWHM) XFEL pulse indicated by the grey line.}
    \label{fig:Ivelocity}
\end{figure}

A one-dimensional cut along the $x$-axis through the momentum distribution in \autoref{fig:alignment}
is shown in \autoref{fig:Ivelocity}. It represents the measured \Iplus-velocity distribution
$v_{\text{\Iplus}}$ in the laboratory frame. The data can be approximated by a Gaussian distribution
with mean $\mu=2700$~m/s and width $\sigma=700$~m/s. Considering momentum conservation, the
distribution of the relative velocities $v_{\text{I--I}}$ of the two iodine atoms is then given by a
Gaussian distribution with $\mu_v=4200$~m/s and $\sigma_v=1090$~m/s. Since a complete velocity
distribution of all ions has not been determined experimentally, this model assumes fragmentation
into \Iplus\ and $\left[\text{C}_7\text{H}_3\text{IN}\right]^{+n}$.\footnote{We note that this model
   contains two simplifications: First, the time for acceleration of the fragments as well as the
   time for ionization, \ie, the finite delay for Auger decay and subsequent charge rearrangement
   after photoabsorption was not taken into account (\ie, set to zero). Therefore, our model
   overestimates the atomic displacements. However, this is partly counteracted by the fact that
   higher charged \Iplusn\ fragments recoil faster than \Iplus\ and hence lead to larger atomic
   displacements, which is not considered, because these momentum distributions of higher charged
   \Iplusn\ fragments were \ not measured.} The resulting velocity distribution of \Iplus fragments
from ionized molecules is
\begin{equation}
   v_\text{I--I}=C\cdot\exp\left(-\frac{(v-\mu_v)^2}{2\sigma_v^2}\right)\label{eq:1}
\end{equation}
with the normalization constant $C$ (such that $\int v_\text{I--I} dv=1$).
This translates into a spatial distribution of I--I distances $s(\Delta{t},d)$ by the substitution
$\liod=v\;\Delta{t}$, with the period $\Delta{t}=t-t_i$ between ionization time $t_i$ and observation
time $t$.

At each time $t$ the probability for photoabsorption and ionization of molecules is
\mbox{$f_\text{ionized}(t)=I_{\text{FEL}}(t)\cdot\sigma_\text{abs}\cdot N /A_0$} with the FEL
intensity $I_{\text{FEL}}(t)$, the photoabsorption cross section $\sigma_\text{abs}$, the number of
molecules $N$, and the interaction area $A_0$. $N/A_0$ can be substituted by $\M\cdot l$ with the
molecular beam density \M\ and the length of the interaction volume in $z$-direction $l$ (see
\autoref{sec:simulation}), hence
\mbox{$f_\text{ionized}(t)=I_{\text{FEL}}(t)\cdot\sigma_\text{abs}\cdot\M\cdot{}l$}. For each time
$t$, the distribution of I--I distances is given as the sum of intact-molecules with distances $d_0$
and the distributions of all previously ionized molecules
\begin{equation}
   s(t, \liod) = F_\text{intact}(t)\cdot{}N\cdot{}s(0,d_0) + \sum\limits_{t_i=0}^{t} s(t-t_i,\liod) \cdot f_\text{ionized}(t_i)
   \label{eq:EffSpatialDistr}
\end{equation}
with the fraction $F_\text{intact}(t)$ of intact molecules at time $t$ and the fraction
$f_\text{ionized}(t_i)$ of molecules ionized at a certain particular time $t_i$ with the property
that \mbox{$\sum_{t_i=0}^t f_\text{ionized}(t_i)=F_\text{ionized}(t)$}, see
\autoref{fig:Ivelocity}~b.

\begin{figure}[t]
    \centering
    \includegraphics[width=\linewidth]{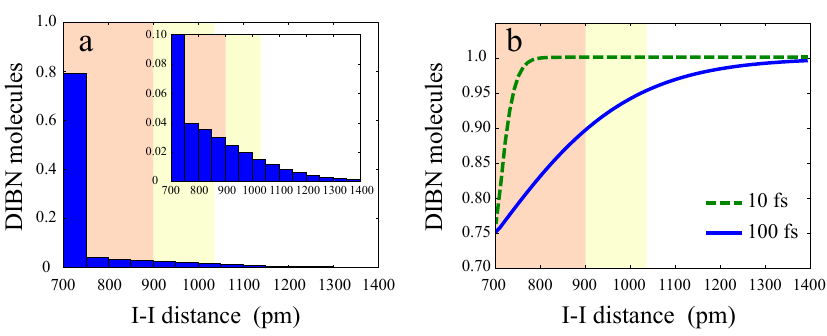}
    \caption{(a) Histogram of  \Sii, visualizing the fraction of molecules in different distance
       intervals, as seen by the 100~fs (FWHM) FEL pulse (blue). (b) The cumulative distribution of
       \Sii\ (for smaller stepsize in I--I-distance). I--I distances that are less than 200~pm (330~pm)
       longer than the 700~pm equilibrium distance are marked by the red (yellow) shaded regions. The
       dashed green graph in (b) is for a theoretical case of a 10~fs pulse.}
    \label{fig:IodineDistanceDistr}%
\end{figure}

The spatial distribution of I--I distances as seen by the FEL pulse is the sum over $s(t,d)$ for all
times, weighted by the instantaneous normalized FEL intensity
$I^\text{norm}_\text{FEL}(t)=I_\text{FEL}(t)/\sum{}I_\text{FEL}(t)$:
\begin{equation}
   S(d)=\sum\limits_{t}s(t,\liod)\cdot{}I^\text{norm}_\text{FEL}(t)
   \label{eq:EffSpatialDistrTotal}
\end{equation}
This distribution is shown in \autoref{fig:IodineDistanceDistr}~a.
The corresponding cumulative distribution of I--I distances is illustrated by the solid blue line in
\autoref{fig:IodineDistanceDistr}~b. The latter gives the amount of molecules with I--I distances
equal to or less than the given distance summed over the entire FEL pulse, \eg, 75~\% of the
elastically scattered photons originate from scattering at intact molecules (\ie, I-I-distance at
equilibrium distance of 700~pm) and another 15~\% (20~\%) of the diffraction signal originates from
scattering of molecules corresponding to I--I distances that are less than 200~pm (330~pm) longer
than the 700~pm equilibrium distance. These distances correspond to the red (yellow) shaded regions
in \autoref{fig:IodineDistanceDistr}. These damaged molecules might contribute to the experimentally
determined elongated I--I distance of 800~pm in the minimum of the \chisqr-fit. However, since the
range of $s$-values (scattering vectors) covered is too small, these effects cannot be fully
resolved in the current experiment with 620~pm wavelength radiation. Further suppressing such
effects on the diffraction pattern could, for instance, be accomplished by using shorter pulses. For
10~fs practically no damage would be observed and even for the same pulse energy $95$~\% of the
molecules would be at equilibrium distance to within 40~pm.

\section{Conclusion and Outlook}
\label{sec:outlook}
We experimentally demonstrated coherent x-ray diffractive imaging of laser-aligned gas-phase samples
of the prototypical complex molecule 2,5-diiodobenzonitrile at the LCLS XFEL. This x-ray diffraction
experiment resembles Young's double slit experiment on the atomic level due to the two-center interference
of the two heavy iodine atoms. We implemented a
state-of-the-art molecular beam setup in the CAMP experimental chamber at the AMO beamline of LCLS,
utilized quantum-state selection of a cold molecular beam, and demonstrated the preparation of a
strongly aligned ensemble of isolated gas-phase molecules. The controlled samples of DIBN were
probed by the x-ray pulses in order to measure x-ray diffraction from these ensembles of aligned
DIBN. Exploiting the high spectral resolution of the pnCCD detectors, we could successfully retrieve
single scattered photons above noise and derive the molecular diffraction patterns from the weak and
noisy signals. On average, $0.2$~photons/shot were recorded on the camera. However, the
angular structures contained in the diffraction patterns are well beyond experimental noise, \ie, we
succeeded to observe the two-center interference of the two heavy iodine atoms in the diffraction
pattern which confirms the observation of a successful diffraction measurement from aligned DIBN.
Even despite the limited resolution, \ie, the long wavelength and the correspondingly limited range
of scattering vectors $s$ recorded, the heavy-atom distance was experimentally obtained and it is
consistent with the computed molecular structure. Future experiments toward
atomic resolution imaging will have to use shorter wavelength and collect diffraction data at higher
resolution.

Our experiment confirms the feasibility of coherent x-ray diffractive imaging of small isolated
gas-phase molecules and hence provides a first step towards single-molecule imaging at atomic
resolution. Our controlled delivery approach is capable to provide three-dimensional alignment and
orientation,\cite{Larsen:PRL85:2470, Nevo:PCCP11:9912, Hansen:JCP139:234313} which would allow the
determination of the 3D molecular structure using a tomographic approach similar to electron
diffraction\cite{Hensley:PRL109:133202} or photoelectron tomography.\cite{Maurer:PRL109:123001}

Envisioned future experiments plan to make use of the unique short pulses of the XFELs in order to
conduct fs pump-probe experiments in order to investigate ultrafast structural dynamics during, \eg,
chemical reactions and open up a new field for experiments in femtochemistry and molecular dynamics.
For the recording of molecular movies of ultrafast dynamics, x rays offer several advantages over
electrons: x-ray pulses do not suffer from space-charge broadening of pulses nor from pump-probe
velocity mismatch.\cite{Williamson:CPL209:199310,Sciaini:RPP74:096101} Hence, x-ray pulses from
XFELs will permit better temporal resolution. Pulses as short as a 2--5~fs are already routinely
created at XFELs,\cite{Ding:PRL102:254801,Ding:PRL109:254802} and attosecond x-ray pulses are
discussed.\cite{Dohlus:PRSTAB14:090702} These short pulses will allow the observation of the fastest
nuclear motion and, moreover, the investigation of ultrafast electron dynamics, such as charge
migration and charge transfer processes in molecular and chemical
processes.\cite{Weinkauf:JPC100:18567, Zewail:JPCA104:5660}

We analyzed how damage effects can be avoided by using short pulses of low fluence at high
repetition rates, which will be available at future XFELs, such as the upcoming European XFEL that
will operate at 27\,000 x-ray pulses/second. Our approach is suitable to study larger molecules
provided moderately dense molecular beams of these samples can be generated. Hence, it should be
applicable for coherent diffractive imaging of isolated biomolecules, as envisioned for a long
time.\cite{Spence:PRL92:198102, Spence:ActaCrystA61:237, Starodub:JCP123:244304, Barty:ARPC64:415}

\section*{Acknowledgements}
We thank Marcus Adolph, Andrew Aquila, Sa\v{s}a Bajt, Carl Caleman, Nicola Coppola, Tjark Delmas,
Holger Fleckenstein, Tais Gorkhover, Lars Gumprecht, Andreas Hartmann, Günter Hauser, Peter Holl,
Andre Hömke, Faton Krasniqi, Gerard Meijer, Robert Moshammer, Christian Reich, Robin Santra, Ilme
Schlichting, Carlo Schmidt, Sebastian Schorb, Joachim Schulz, Heike Soltau, John C.\ H.\ Spence,
Lothar Strüder, Joachim Ullrich, Marc J.\ J.\ Vrakking, and Cornelia Wunderer for help in preparing
or performing the measurements.

Parts of this research were carried out at the Linac Coherent Light Source (LCLS) at the SLAC
National Accelerator Laboratory. LCLS is an Office of Science User Facility operated for the
U.~S.~Department of Energy Office of Science by Stanford University. We acknowledge the Max Planck
Society for funding the development and operation of the CAMP instrument within the ASG at CFEL.
A.Ro. acknowledges the research program of the "Stichting voor Fundamenteel Onderzoek der Materie",
which is financially supported by the "Nederlandse organisatie voor Wetenschappelijk Onderzoek".
H.~S.\ acknowledges support from the Carlsberg Foundation. P.~J.\ acknowledges support from the
Swedish Research Council and the Swedish Foundation for Strategic Research. H.N.C.\ acknowledges
NSF STC award 1231306. A.Ru.\ acknowledges support from the Chemical Sciences, Geosciences, and
Biosciences Division, Office of Basic Energy Sciences, Office of Science, US Department of Energy.
D.~R.\ acknowledges support from the Helmholtz Gemeinschaft through the Young Investigator Program.
This work has been supported by the excellence cluster ``The Hamburg Center for Ultrafast Imaging
-- Structure, Dynamics and Control of Matter at the Atomic Scale'' of the Deutsche Forschungsgemeinschaft.

\theendnotes

\newpage
\appendix
\renewcommand{\thesection}{Appendix~\Alph{section}}
\section{Data acquisition and conditioning of x-ray diffraction data}
\label{sec:appendix:daq}
X-ray diffraction data was recorded by the pnCCD photon detectors with the LCLS operating at $60$~Hz.
The YAG was operating at $30$~Hz, hence single-shot YAG and NoYAG data was recorded in an alternating
manner. The YAG and LCLS laser were propagating collinearly (see \autoref{fig:setup}) which resulted
in severe background levels from the YAG on the pnCCD despite the filters. This background as well as
camera artifacts, known from dark frame measurements, were subtracted from the single shot data. The
necessary single-photon counting required operation of the pnCCD cameras at the highest possible gain
in order to give a good separation of 2~keV and optical and NIR photons (the latter from  the YAG).
Spectroscopic discrimination of rare events, \ie, single 2~keV x-ray photons, could be performed due
to the high energy resolution of the pnCCD camera.\cite{Strueder:NIMA614:483} In this chapter we
describe the steps necessary to correct the single-shot diffraction data for all artifacts and
backgrounds. All processing of the data was performed using the CFEL-ASG Software Suite
(CASS).\cite{Foucar:CPC183:2207}

\begin{figure}[t]
   \centering%
   \includegraphics[width=1.00\textwidth]{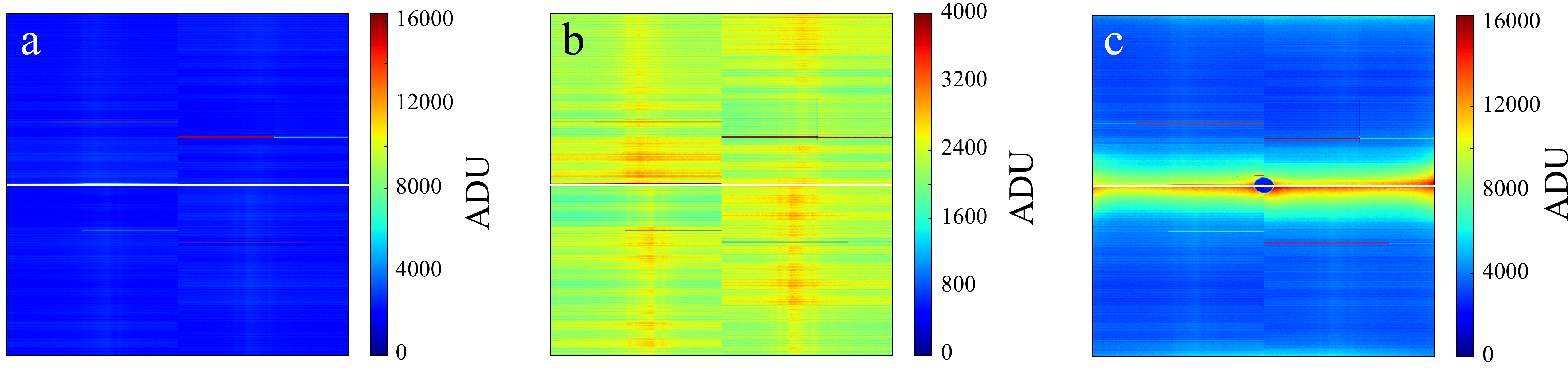}%
   \caption{Single shot raw data frames of an example dataset for the NoYAG case (a,b) and the YAG
      case (c).}%
   \label{fig:rawdata}%
\end{figure}
\autoref{fig:rawdata} shows typical examples of single shot raw data frames for both panels of the
pnCCD camera for (a,~b) NoYAG and (c) YAG, which contain many artifacts. The measured pnCCD signals
are given in ADU (analog-to-digital unit). The most significant difference between NoYAG and
YAG data is the region of partly saturated signal at the inner edges of the two pnCCD
panels in the YAG case. These signals are
based on imperfect shielding of both pnCCD panels from near-infrared (NIR) photons especially at
their respective edges.\footnote{The charge created by a single YAG photon is less than a 1/500th of
   the charge created by a single 2~keV x-ray photon. However, due to the high YAG intensity, many
   YAG photons pile up in a single pixel, especially in the regions not shielded thoroughly by the
   filters.} This contribution to the experimental background is referred to as ``YAG background''.
Furthermore, the single shot data contains pnCCD based artifacts such as offset- and gain
variations, ``hot pixels'' or even ``hot rows/channels'', and time-dependent readout fluctuations
called ``common mode'' (during read out of the pnCCDs, charges are shifted towards the ASIC along
the horizontal direction). The pnCCD consist of 16 CAMEX modules.\footnote{For a description of
the CAMEX modules, see references \onlinecite{Strueder:NIMA614:483, Hartmann:NSSCR:2590}.}
The pnCCD-based artifacts and distinct offset within the 16 CAMEX modules become more obvious when
zooming into the colorscale, see \autoref{fig:rawdata} (b).\footnote{Although there is a
channel-specific offset and gain variation, all channels within the same CAMEX have similar gain
and the difference of the channel-specific gain between distinct CAMEX is more pronounced than the
gain variation within one CAMEX.}

In our experiment, scattering from ensembles of isolated molecules is very weak; in particular the
probability for two or more x-ray photons scattered to the same pixel on the detector within the
same single shot is negligible small. Therefore, single x-ray photon hits could be found by spectroscopic,
\ie, energy-dependent discrimination of single-shot data which was corrected for all pnCCD artifacts and the
YAG background. The measured ADU value is proportional to the energy and a single 2~keV x-ray photon
corresponds to a value of $\approx\!2600$~ADU.

\begin{figure}[tb]
   \centering
   \includegraphics[width=0.55\textwidth]{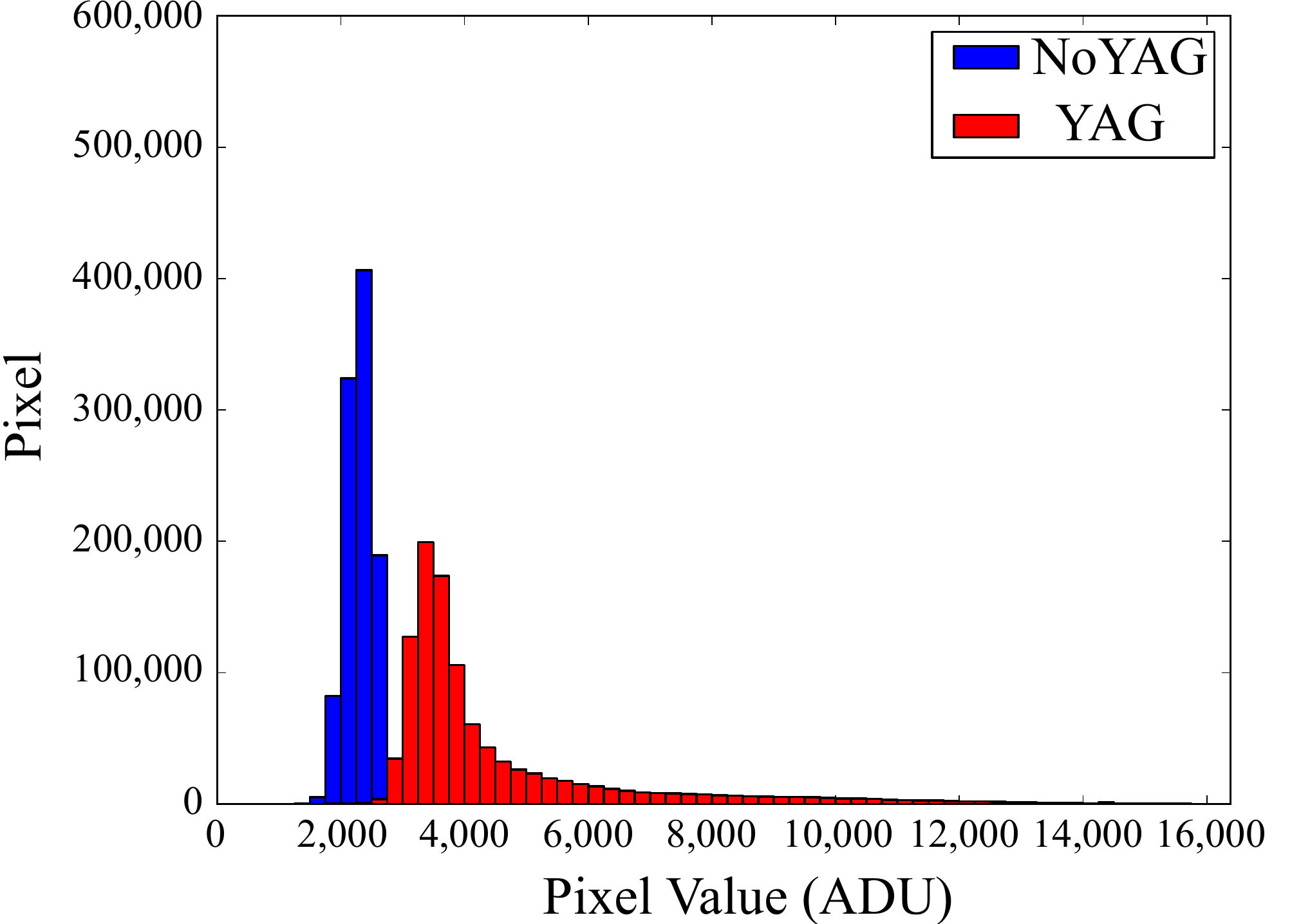}
   \caption{Spectra for the single shot data frames given in \autoref{fig:rawdata}~a, c; see text
      for details.}
   \label{fig:spectrum}
\end{figure}
\autoref{fig:spectrum} shows histograms for the $1024\times 1024$ pnCCD values of the single shot
data frames given in \autoref{fig:rawdata} (a--c). There is a constant offset of $\approx\!2400$~ADU
in most pixels and in both cases (YAG and NoYAG). In the YAG case, in addition to the pnCCD-based offset,
there is the huge background at the inner (and outer edges) of the two pnCCD panels, resulting in a shift of
the spectrum towards higher values.

\begin{figure}[b]
    \centering
    \includegraphics[width=0.80\textwidth]{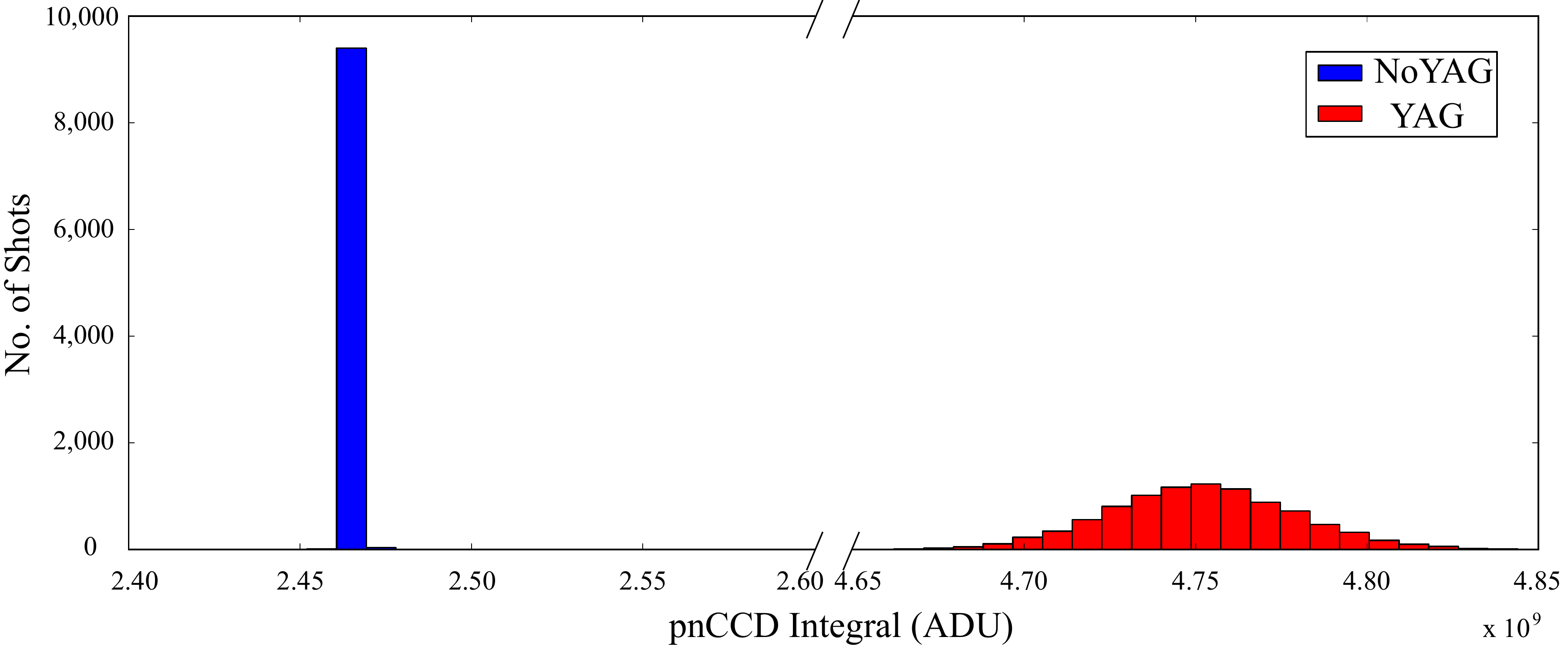}
    \caption{Histogram of the total integrated value of individual YAG/NoYAG data for an example
       dataset containing 9451 shots with YAG off (NoYAG) and 9449 shots with YAG on (YAG).}
    \label{fig:CCDintegral_RawData}
\end{figure}
The YAG background was utilized to reliably distinguish single-shot YAG from single-shot NoYAG data.
\autoref{fig:CCDintegral_RawData} shows a histogram of the integrated pnCCD signals for single
YAG/NoYAG shots for an example dataset, illustrating the clear separation of YAG and NoYAG shots.
The variation in the YAG case is due to the YAG intensity, varying on a shot-to-shot level.

First during the data conditioning process, single-shot YAG and NoYAG data was separated based on
the integrated pnCCD signals. Then, the
data was corrected for offset by subtracting an offset map, the latter obtained from averaging
single shot pnCCD data under ``dark'' conditions. The common mode was corrected for by subtracting
the median value along each vertical row from this row, separately for the upper and lower pnCCD
panel.
\begin{figure}[tb]
    \centering
    \includegraphics[width=0.85\textwidth]{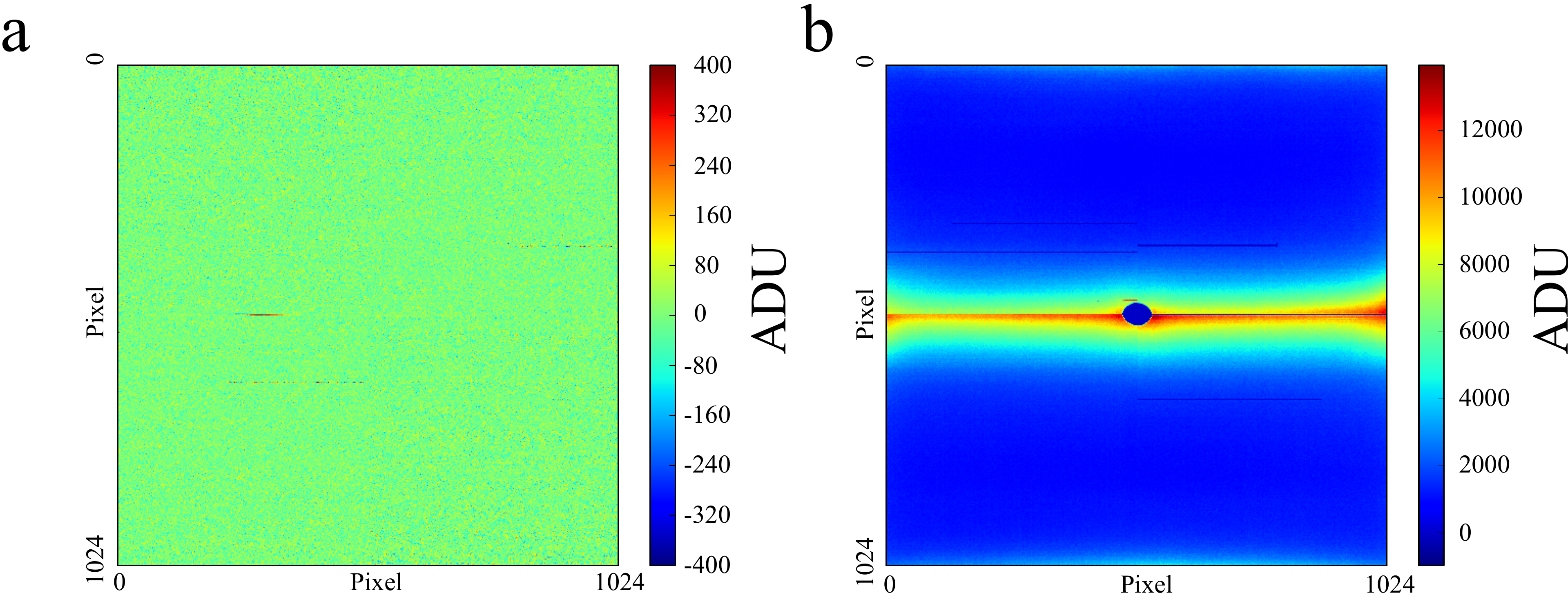}
    \caption{Single shot pnCCD data frames, corrected for channel-dependent offset and common mode.}
    \label{fig:OffsetCommonMode}
\end{figure}
\autoref{fig:OffsetCommonMode} shows the resulting frames for the YAG and NoYAG case. The
channel-specific offset variation was successfully corrected for. The NoYAG data is close to 0 value
for almost every pixel while the YAG data still contains the severe background from the YAG.

The YAG background scattering was corrected for by subtracting a averaged YAG data frame, scaled
to match the total intensity of the particular individual single-shot data frame, from the
individual single-shot YAG frame. This method works reliable since the total YAG intensity is varying
on a shot-to-shot level but the spatial distribution of the YAG on the pnCCD is independent of a
certain shot (\ie, it can be scaled by a single number).

\begin{figure}[t]
    \centering
    \includegraphics[width=1.00\textwidth]{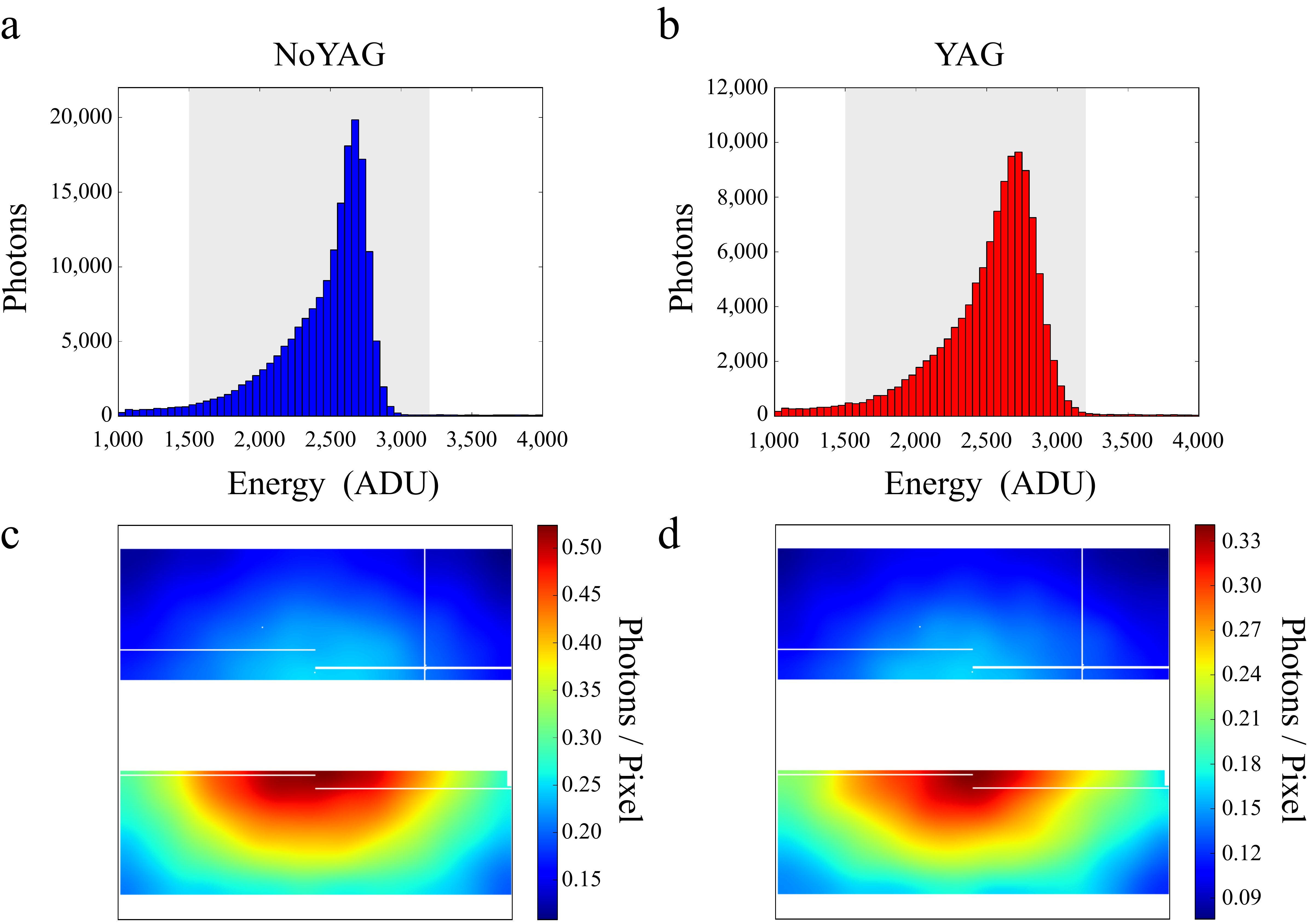}
    \caption{Spectra of the hits for NoYAG (a) and YAG (b) for hits made up out of 1--2~pixels;
        spatial intensity distributions \IYAG, \INoYAG\ of these hits in the energy interval
        1500--3200 ADU (c,~d), \ie, the ``diffraction patterns''. The raw data was convolved with
        a gaussian kernel.}
    \label{fig:Finished_Pixels1-2}
\end{figure}

As a result from the steps mentioned above, the single-shot data frames were corrected for
all backgrounds and artifacts except rare events such as single 2~keV x-ray photons scattered from
the molecular sample. These photons, at $\approx2600$~ADU, were found by thresholding the
background-corrected data frames and considering the charge spread of the x-ray photons: a 2~keV
photon absorbed
in the pnCCD creates a charge cloud which can cross the barrier of a single pixel and hence can give
signals in two (or more) adjacent pixels. At 2~keV, almost all photon hits are single- or
double-pixel hits (the latter is the case in which the charge cloud diffuses into a single
neighboring pixel adjacent to the pixel where the photon is initially absorbed). This is justified
by the experimental results, where 64~$\%$ of all x-ray attributed hits are single-pixel hits,
35~$\%$ are double-pixel hits while $<1~\%$ make up for the rest. These photon hits were found by
thresholding the background-corrected single-shot YAG and NoYAG data frames and combining adjacent pixels
exceeding the threshold of 500~ADU. The x-ray hits found by this procedure were written to a list
containing the coordinates, ADU value, and number of pixels the hit was combined from. By
limiting the number of pixels a hit can be made of to six, rare events such as high energy particles
impinging on the detector were neglected. Then, corrections for channel-dependent gain and
charge-transfer-efficiency were applied to the energy values of the photon hits (although these
corrections didn't affect the spatial distribution and also have almost no effect on the spectral
distribution of the hits as well). Photon hits for certain regions of pixels were always neglected.
This included ``hot pixel'' regions as well as the parts of the pnCCD that were (completely or nearly)
saturated by YAG photons. The latter regions showed a high fluctuation of signal and, therefore,
could not be thresholded successfully.

\autoref{fig:Finished_Pixels1-2}~a,~b show spectra of all photon hits from the NoYAG (a) and YAG (b)
data. The spectrum is peaks at 2600~ADU, thereby matching expectations. The width of the peak can be
attributed the energy resolution of the pnCCDs, the photon energy jitter of LCLS, and to the event
recombination of double pixel hits (the latter being the major contribution to the broadening of the
spectrum).

In the energy interval 1500--3200~ADU there are 172\,499 photons for the NoYAG and 111\,560 photons
for the YAG data which are used for data analysis. The data was obtained from 842\,722 shots (NoYAG)
and 563\,453 shots (YAG) respectively, hence the average hit rate on the whole pnCCD detector was
0.204 (0.197) photons/shot for the NoYAG (YAG) data. The spatial distribution of these photon hits,
\ie, the diffraction patterns \INoYAG\ and \IYAG, are shown in \autoref{fig:Finished_Pixels1-2}~c,~d
and are analyzed as described in \autoref{sec:results}.

\providecommand*{\mcitethebibliography}{\thebibliography}
\csname @ifundefined\endcsname{endmcitethebibliography}
{\let\endmcitethebibliography\endthebibliography}{}

\end{document}